\documentclass[12pt,a4paper]{report}

\usepackage[BCOR=1cm]{typearea}
\usepackage[german, USenglish]{babel}
\usepackage{tocbibind}
\usepackage{parskip}
\usepackage[utf8]{inputenc}
\usepackage{a4wide}
\usepackage{makeidx}
\usepackage{url}
\usepackage{epsfig}
\usepackage{epstopdf}
\usepackage{doc}
\usepackage{graphicx}
\usepackage{amssymb}
\usepackage{amsmath}
\usepackage{verbatim}	%for multi line comments
\usepackage{tikz}
\usepackage{diagbox}

\usepackage{url}

\usepackage{array}

\newcommand{\setlang}[1]{\selectlanguage{#1}\nonfrenchspacing}

\begin{document}

% TITLE:

\thispagestyle{empty}
\newpage

\vspace{5cm}
\begin{center}
    \epsfxsize=4cm
    \epsfbox{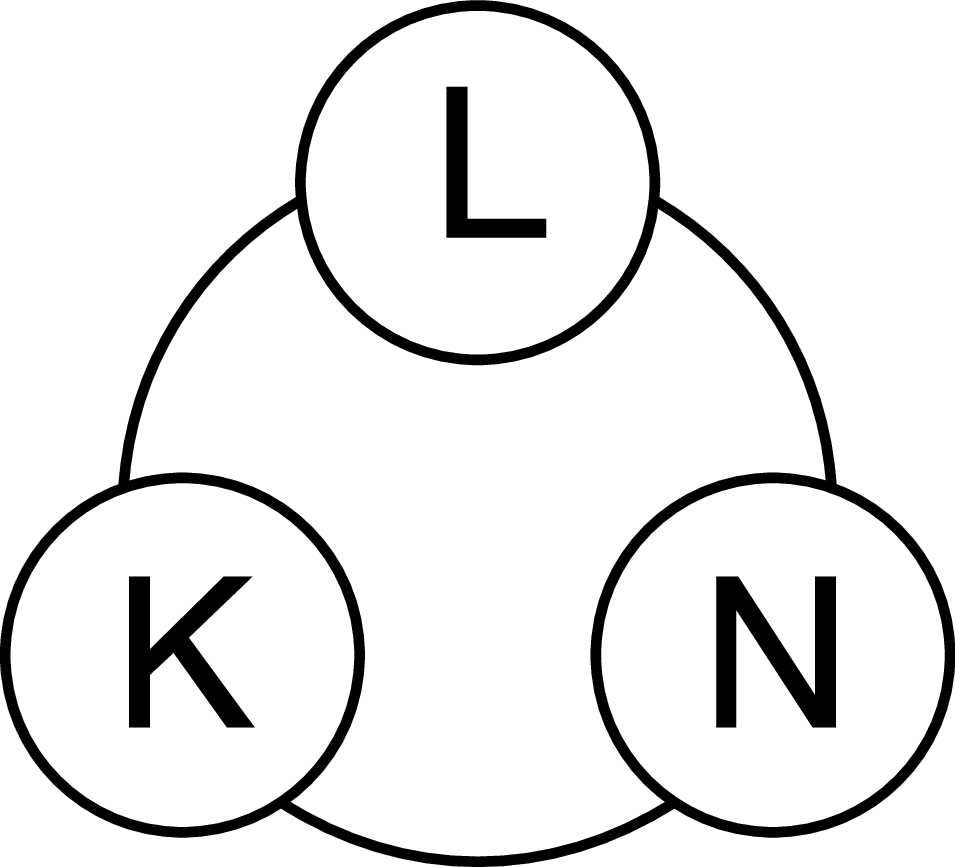}
\end{center}

\parbox{15cm}{\begin{center} {\sf\bf 
                               \Large  Technische Universität München
                                \smallskip

                               \Large Lehrstuhl für Kommunikationsnetze
                               \smallskip
                              }

                              {\sf \large Prof. Dr.-Ing. Wolfgang Kellerer} 
              \end{center}}  %&

\vspace{4cm}

\begin{center}
        {\bf\Huge Bachelor's Thesis} % Studienarbeit, Interdisziplinäres Projekt ‘
\end{center}

\begin{center}
        \settowidth{\baselineskip}{0.4cm}
        {\LARGE 
        Development and Justification of a Physical Layer Model Based on Monitoring Data for Quantum Key Distribution
        }
\end{center}

\vfill         
{\settowidth{\baselineskip}{0.2cm}
\large\begin{tabular}[l]{ll}
Author: & Haiden, Gian-Luca\\
Address: & Schlattweg 5\\
         & 6922 Wolfurt\\
         & Austria\\
Matriculation Number: & 03742781\\
Supervisor: & Wenning, Mario\\
Begin: & 22. January 2024\\
End: & 10. June 2024
\end{tabular}}

% Switch to roman numbering
%\pagenumbering{roman}

%%%%%%%%%%%%%%%%%%%%%%%%%%%%%%%%%%%%%%%%%%%%%%%%%%%%%%%%%%%%

% MAIN PART:
% Independence and License statements
\thispagestyle{plain}

\section*{Preface} 

This bachelor's thesis is the final component of the Bachelor of Electrical Engineering and Information Technology at TUM, the Technical University of Munich. 

I would like to thank my supervisor, Mario Wenning, for his valuable insights and guidance throughout this and the development process.

\vspace*{1cm}
With my signature below, I assert that the work in this thesis has been composed by myself independently and no source materials or aids other than those mentioned in the thesis have been used.

\vspace{1.5cm}

\hspace{1cm}\begin{tabular}{ccc}
\vspace{-0.3cm}München, 10.06.2024	&\hspace{4cm} 		& \includegraphics[width=0.3\textwidth]{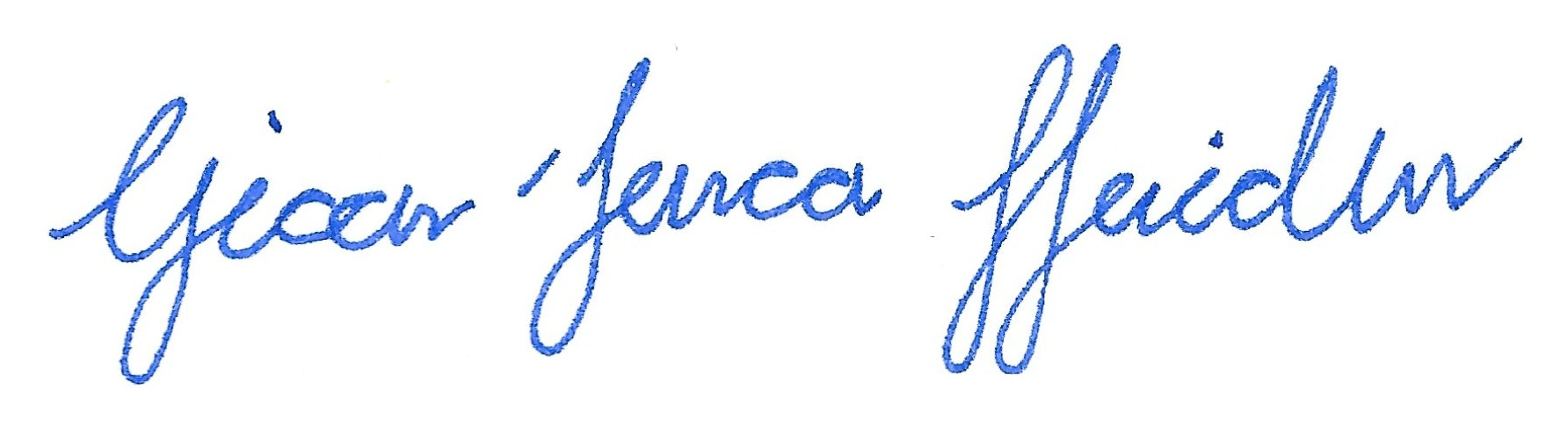}	\\
\rule{4.5cm}{0.4pt}					&					&\rule{4.5cm}{0.4pt}	\\
Place, Date							&					& Signature			
\end{tabular}

\vspace{3cm}
This work is licensed under the Creative Commons Attribution 3.0 Germany License. To view a copy of the license, visit http://creativecommons.org/licenses/by/3.0/de\\

Or\\

Send a letter to Creative Commons, 171 Second Street, Suite 300, San Francisco, California 94105, USA.

\vspace{1.5cm}

\hspace{1cm}\begin{tabular}{ccc}
\vspace{-0.3cm}München, 10.06.2024 	&\hspace{4cm} 		& \includegraphics[width=0.3\textwidth]{graphics/sign.png}	\\
\rule{4.5cm}{0.4pt}					&					&\rule{4.5cm}{0.4pt}\\
Place, Date							&					& Signature	
\end{tabular}

% German abstract:
\setlang{german}
\thispagestyle{plain}

\section*{Kurzfassung} 
Quantum Key Distribution (QKD) ist eine neue, vielversprechende Technologie für zukunftssichere Kommunikationssysteme. Anders als ein konventioneller Schlüsselaustausch mittels RSA, welcher theoretisch nicht genug Sicherheit gegenüber Quantencomputern bietet \cite{shor1999algorithm}, eröffnet QKD die Möglichkeit einer weitaus besseren Sicherheit auf Grundlage der Quantenmechanik \cite{nurhadi18qkdsurvey}. Trotz der technischen Raffinesse und der mittlerweile auch größeren Verfügbarkeit von QKD Systemen, ist die Dokumentation dieser Systeme meist unzureichend, wodurch sich die Funktionalität nicht im Detail erschließen lässt. Diese Bachelorarbeit befasst sich mit den praktischen Mängeln die reale QKD Systeme aufweisen wie z.\,B. ein instabiler Betrieb infolge von schwankenden Schlüssel-Übertragungsraten (SKR = Secret Key Rate). Durch die theoretische Analyse von Messdaten eines QKD Systems in Polen konnten erste Leitungsabschätzungen, zu dieser Form der Analyse und zur Machbarkeit einer SKR Vorhersage, gemacht werden. Es wurde ein Ansatz gewählt, welcher mithilfe von Maschinellem Lernen (ML ,engl: Machine Learning) die SKR, und somit das Verhalten eines QKD Systems, zuverlässig vorhersagen kann.

Die beschriebenen Methodiken konzentrieren sich vorallem auf die Entwicklung von einem theoretischen QKD Modell und der Implementierung von ML Modellen mittels Keras, einem speziell dafür zugeschnittenen Tool von Tensorflow \cite{tensorflow2015-whitepaper}. Eine wichtige gewonnene Erkenntnis ist dabei, dass das theoretische Modell zwar grundlegende Einblicken in QKD Systeme gibt, sich jedoch der ML Ansatz deutlich besser zur Vorhersage der SKR eignet, aufgrund der deutlich höheren Genauigkeit. Diese Arbeit hebt die Grenzen von theoretischen QKD Modellen hervor und unterstreicht die Relevanz von ML Modellen angewendet auf QKD in der Praxis.

Weiterführende Forschung sollte sich im Speziellen darauf fokussieren ein Modell für die Bitübertragungsschicht zu entwickeln, welches eine langfristige Vorhersage der SKR ermöglicht. Mithilfe eines solchen Modells könnte verhindert werden, dass Schlüssel aufgebraucht werden und nicht mehr zur Verfügung stehen, falls die SKR stark abfallen sollte. Zusammengefasst werden in dieser Bachelorarbeit fundamentale Ansätze entwickelt, welche ML für die Prognose des Verhaltes von QKD Systemen nutzt und damit den Weg für weitergehende Forschung zur langfristingen Vorhersage der SKR ebnet.
% English abstract:
\setlang{USenglish}
\thispagestyle{plain}

\section*{Abstract}
Quantum Key Distribution (QKD) is a promising technique for ensuring long-term security in communication systems. Unlike conventional key exchange methods like RSA, which quantum computers could theoretically break \cite{shor1999algorithm}, QKD offers enhanced security based on quantum mechanics \cite{nurhadi18qkdsurvey}. Despite its maturity and commercial availability, QKD devices often have undisclosed implementations and are tamper-protected. This thesis addresses the practical imperfections of QKD systems, such as low and fluctuating Secret Key Rates (SKR) and unstable performance. By applying theoretical SKR derivations to measurement data from a QKD system in Poland, we gain insights into current system performance and develop machine learning (ML) models to predict system behavior.

Our methodologies include creating a theoretical QKD model \cite{nurhadi18qkdsurvey} and implementing ML models using tools like Keras (TensorFlow \cite{tensorflow2015-whitepaper}). Key findings reveal that while theoretical models offer foundational insights, ML models provide superior accuracy in forecasting QKD system performance, adapting to environmental and operational parameters. This thesis highlights the limitations of theoretical models and underscores the practical relevance of ML models for QKD systems. 

Future research should focus on developing a comprehensive physical layer model capable of doing long-term forcasting of the SKR. Such a model could prevent an encryption system form running out of keys if the SKR drops significantly. In summary, this thesis establishes a foundational approach for using ML models to predict QKD system performance, paving the way for future advancements in SKR long-term predictions.
% Table of contents:
\tableofcontents  

% Switch to arabic numbering
%\clearpage
%\pagenumbering{arabic}

% Introduction (Einleitung):
\chapter{Introduction}

\section{Motivation}
Quantum Key Distribution (QKD) is a promising technique for ensuring long-term security in communication systems.
Unlike conventional key exchange methods like RSA, which rely on complex mathematical functions resistant to classical computers, quantum computers can in theory efficiently break these algorithms, posing a future security risk \cite{shor1999algorithm}. However, quantum computers on the necessary scale are not yet available. Over recent years, QKD has grown in maturity and devices are now commercially available. Even still, the details concerning the implementation of these devices is often not publicly disclosed and is typically tamper-protected.

Nevertheless, having access to monitoring data from a meshed QKD network allows us to gain valuable insights into the performance and limitations of current QKD systems by applying theoretical models of the Secret Key Rate (SKR) to measured data. This allows for an examination of the plausibility of SKR with regard to measured physical parameters to create a deeper understanding of the system behavior. Additionally, fitting curve parameters to the measured data can enhance modeling accuracy. Lastly, developing machine learning (ML) models enables further generalization as well as the estimation of unknown links.

\section{Objectives}
Quantum key distribution (QKD) is a new key exchange technology that offers to communicate securely, depending on quantum mechanics laws \cite{nurhadi18qkdsurvey}.\\
While QKD is secure from an information-theoretic perspective, practical implementation challenges and imperfections can complicate being able to achieve a high level of security in real-world applications.
Encrypted communication still requires an encryption algorithm, though this will not be discussed in this thesis. QKD is solely responsible for the secure exchange of cryptographic keys.
While theoretical derivations for QKD exist, practical use cases reveal significant implementation challenges and errors. Although there are already existing QKD networks \cite{hubel2023deployed}, their discussions on difficulties are not profound enough. Unfortunately, not even the manufacturers of such QKD devices (Cerberis\textsuperscript{3} from ID Quantic \cite{idq2019cerberis3}) disclose explicit knowledge about how their device acts and tunes parameters; hence, this information cannot be found in the datasheets. 
Non-ideal photon sources and detectors (efficiency), fiber attenuation, dark count, the average photon number per pulse and losses in other components have a significant influence on the key exchanges. These imperfections lead to a fluctuation of the SKR.

This thesis aims to analyze the fluctuation of the SKR. In order to achieve this, data from a build-up system in Poland is used. The goal is to analyze this data and develop a ML model that accurately represents this system. This model shall then be able to predict the behavior of another QKD system.

\section{Thesis Structure}
The thesis is structured as follows: Following the introductory chapter, there is an explanation of the background of QKD. This background provides insights to QKD and a conceptional overview of QKD as well as further exploring the future necessity of QKD and addresses the specific research gap the thesis aims to resolve.

The methodology chapter states and explains methods that are used in the implementation of the data preprocessing and model generation.
It describes what type of QKD system is used. It states which data formats are used, which parameters must be monitored, and how the model is generated.\\
The implementation chapter outlines the actual implementation of the data preprocessing, the statistical analysis, the process of the model building, and the final ML model implementation.\\
In the results chapter, the results are evaluated and discussed. \\
Finally, there is a conclusion and an outlook that explains what essential discoveries were made during this study and what research efforts are needed to further improve operating QKD systems in meshed networks.

% Text Body (Hauptteil)
% Could have multiple chaper-files, e.g.:
\chapter{Background}

This chapter provides the essential background necessary to understand the contributions and impact of this thesis. We begin with a technical overview of network security, introducing key terms and definitions critical to the field. 
Next, we examine the weaknesses of classical encryption methods, highlighting the need for more secure communication systems. As a solution, we introduce quantum key distribution (QKD) and explain its principles and advantages. Within QKD, we focus on the Coherent-One-Way (COW) protocol, detailing its unique features.
The chapter concludes with a statement on the objectives of the thesis, outlining the primary goals and setting the agenda for the subsequent chapters. This comprehensive background sets the stage for understanding of the research and the specific problems it aims to address.

\section{Technical Background}
\subsection{Network Security}

\begin{figure}[h]%[h!]
  \begin{center}
    \includegraphics[width=0.9\textwidth]{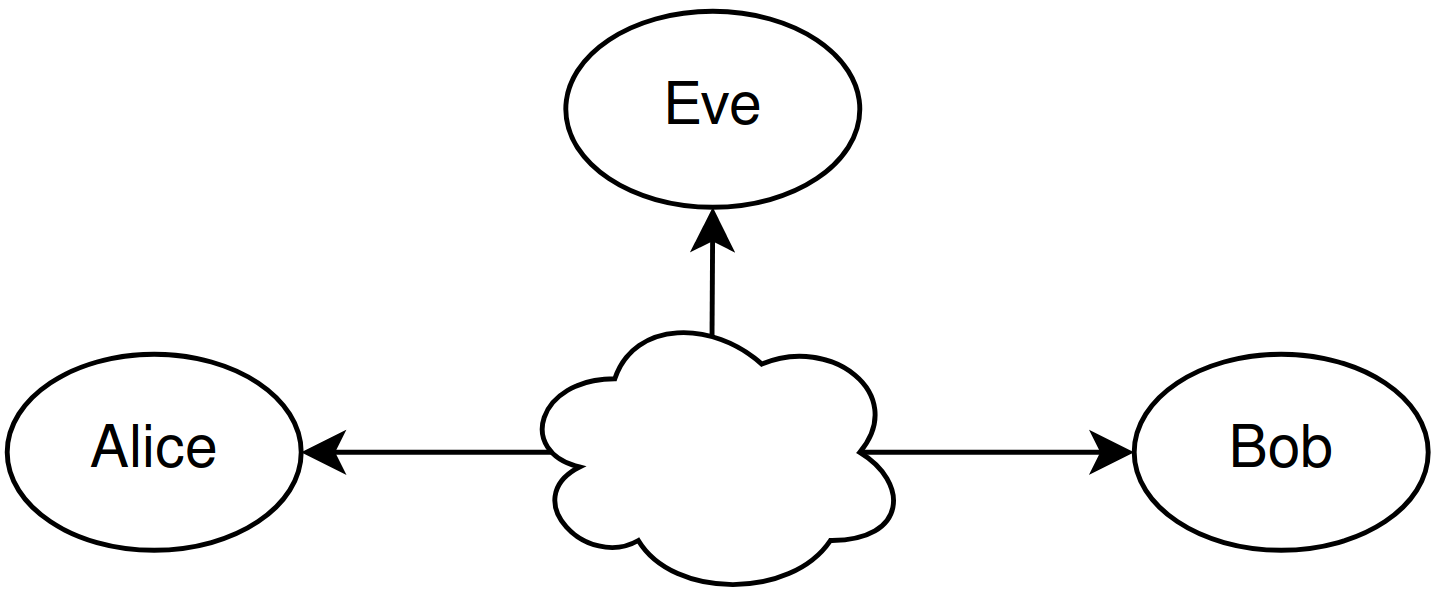}
    \caption{Illustration of the relationship between Alice, Bob and Eve}
    \label{fig:alicebobeve}
  \end{center}
\end{figure}

~Figure~\ref{fig:alicebobeve} describes a communication between Alice, the legitimate sender, Bob, the legitimate receiver and Eve, a non-legitimate eavesdropper. Secure communication aims to transmit information between Alice and Bob without Eve being able to intercept the messages sent.
 Classical encryption and authentication are used to secure the connection between Alice and Bob. 

\subsection{Terms and Definitions}
This section provides definitions of the terms used in the scope of this thesis.
In this thesis, we refer to QKD as a key exchange process and do not restrict the use of keys to particular encryption algorithms.
The term 'classical encryption' in this thesis refers to all conventional encryption technologies that are used nowadays on a general basis. The Rivest–Shamir–Adleman (RSA) algorithm is an example of a classical encryption algorithm.

\subsubsection*{Entropy}
Entropy, as introduced in \cite{shannon1948mathematical}, quantifies the average level of uncertainty or unpredictability inherent in a set of messages. This measure is essential for understanding the limits of data compression and the capacity of communication channels. Shannon defined the entropy $H$ for a discrete random variable $X$ with a set of possible outcomes $\{x_1, x_2, \ldots, x_n\}$ and an associated probability mass function $P(X)$. The formula used to calculate entropy is:
\begin{equation}
    \begin{split}
        \label{eq:entropy}
		H(X) &= -\sum_{i=1}^n P(x_i) \log_2 P(x_i)
    \end{split}
\end{equation}
In this equation, $P(x_i)$ represents the probability of each outcome $x_i$, and the base-2 logarithm is used to yield a result measured in bits.

\subsubsection*{Mutual Information}
Mutual information quantifies the amount of information shared between two variables. In this thesis, it represents the information shared between Alice and Bob ($I_{ab}$) and between Alice and Eve ($I_{ae}$). It measures how the more that is known about one variable, the more uncertainty of another variable is reduced. Formally, mutual information between two variables \( X \) and \( Y \) is defined as:
\begin{equation}
    \begin{split}
        \label{eq:mutualinf}
        I(X; Y) &= H(X) + H(Y) - H(X,Y)\\
        %I(X; Y) &= \sum_{x \in X, y \in Y} p(x, y) \log \left(\frac{p(x, y)}{p(x)p(y)}\right)\\
    \end{split}
\end{equation}
where $H(X)$ is the entropy of $X$, $H(Y)$ is the entropy of $Y$ and $H(X,Y)$ is the joint entropy of $X$ and $Y$. 
%In the alternative quation, where the entropy is already part of the equation, \( p(x,y) \) is the joint probability distribution of \( X \) and \( Y \), and \( p(x) \) and \( p(y) \) are their marginal distributions. 
Mutual information is introduced under the name rate of transmission in \cite{shannon1948mathematical}.

\subsection{Classical Encryption and its Weaknesses}
Encryption technology, pivotal for securing everything from emails to bank accounts, utilizes two main types: symmetric and asymmetric encryption. Symmetric encryption, such as the Advanced Encryption Standard (AES), requires both parties to possess a shared secret key for encrypting and decrypting messages. The key challenge with symmetric encryption is the secure exchange of its key. If intercepted during transmission, an eavesdropper can access all encrypted communications. Asymmetric encryption, like the RSA algorithm, has traditionally been employed to safely exchange these symmetric keys. RSA uses a pair of keys: A public key that is shared openly and a private key kept secret by the recipient. The security of RSA hinges on the complex task of factoring large prime numbers - a task that cannot be managed by conventional computers but that can be efficiently calculated by a quantum computer. In 1994, Peter Shor developed an algorithm that could efficiently factor these numbers using a quantum computer, posing a significant threat to RSA's security \cite{shor1999algorithm}.\\
QKD offers a revolutionary approach to address these quantum vulnerabilities. QKD uses the principles of quantum mechanics to enable the generation and sharing of a secret key between parties with security. It ensures that any attempt to eavesdrop on the key exchange is minimized. This capability of QKD to securely manage key distribution - without the weaknesses of asymmetric methods like RSA - positions it as an optimal solution for future-proof secure communications, particularly in environments threatened by quantum computing advancements. The One-Time-Pad (OTP) is another symmetric encryption standard that works with QKD. However, the downside is that the key always has to be the same size as the message, which makes encrypted communication highly dependent on the speed of the key generation and transmission. Therefore, the key transmission in QKD is still a bottleneck, which is why AES (AES-256 has been proven quantum secure) is also used in combination with QKD.

\subsection{QKD as a more Secure Solution}	% to these Weaknesses}

This section intends to present QKD as a solution to the weaknesses of classical encryption discussed earlier and explore in further detail the already-mentioned advancements.
QKD can be confused with quantum encryption or quantum communication, which are terms which may be misleading. QKD is not encrypting any quantum bits. Instead, QKD is only involved in the key generation and exchange process. Therefore, it uses quantum states which represent 0s or 1s. These states are then securely transferred from Alice to Bob.\\
%To explain how the security of QKD works, we must first introduce the concept of conjugated bases and eigenstates. Eigenstates $\Psi$ of an operator $\hat{R}|\Psi \rangle = \lambda |\Psi \rangle$ form an orthonormal basis (ONB) of the Hilbertspace $ H : \langle \Psi_i | \Psi_j \rangle = \delta_{ij} $ with the Kronecker delta $\delta$. A photons can be put into eigenstates which are quantum state used used to define a MB.
A measurement basis (MB) is an orthonormal basis (ONB) that is used to measure a photon's quantum state $\Psi$. If $\Psi$ is an eigenstate (ES) in the MB, then the outcome of the measurement is its current state with a likelihood of 100\%. However, if $\Psi$ is not an ES in the MB, then the result will be one of the ES in the MB with a likelihood of 50\% for each ES \cite{bennett1984quantum}.
Typical MBs are $\sigma_x$ and $\sigma_z$, which are conjugate to each other. Conjugated bases are MBs for which the uncertainty is maximal for the ES of the other bases. In other words, this means that measurements for ES in a conjugated basis are maximally random, which is part of the key concept for QKD's security. Another additional factor is that after a measurement, the quantum system is in an ES of its applied measurement operator, which means that a measurement changes a system's state \cite{bennett1984quantum}. This usually happens if $\Psi$ is measured in its conjugated basis.
\begin{figure}%[h]%[h!]
	\begin{center}
		\begin{tikzpicture}
    		% Nodes
    		\node (A) at (0,0.5) 		{Sender A};
    		\node (Alice) at (0,0) 		{\;\;(Alice)\;\;};
    		\node (B) at (8,0.5) 		{Receiver B};
    		\node (Bob) at (8,0) 		{\;\;\;\;(Bob)\;\;\;\;};
    		% Edges
    		\draw[->] (A) -- (B) node[midway, above] {Quantum Channel};
    		\draw[<->] (Alice) -- (Bob) node[midway, below] {bidirectional classical Channel};
			\end{tikzpicture}
    	\caption{Conceptional setup for BB84}
    	\label{fig:bb84}
	\end{center}
\end{figure}

\subsubsection*{BB84}
Having explained the basic terminologies, we will now explain the QKD concept with the famous BB84 protocol \cite{bennett1984quantum}. Figure~\ref{fig:bb84} is used to illustrate the protocol. BB84 uses 2 channels: a quantum channel, which is only used for sending the photons held in their quantum states from Alice to Bob, and the bidirectional classical channel, which is used for the transmission of the MB. Before Alice and Bob can start comunicating, they have to agree upon 2 bases to use. A common choice is to use polarization states for the photons. Therefore, the basis could be rectilinear (horizontal and vertical) polarization for $\sigma_z$ and diagonal (+45° and antidiagonal 135°=-45°) polarization for $\sigma_x$. The 4 states are then $\sigma_z: |H\rangle, |V\rangle$ and $\sigma_x: |+\rangle, |-\rangle$. Each state represents either a logical `0' or `1'. For our example, we use a logical `0' for $|H\rangle$ and $|+\rangle$ and a logical `1' for the states $|V\rangle$ and $|-\rangle$.\\

The BB84 protocol works with the following steps \cite{bennett1984quantum}:

\begin{enumerate}
	\item \textbf{Preparation, transmission and measurement of photons:} Alice selects a random state $\Psi$ out of the 4 choices from the 2 conjugated bases and sends $\Psi$ to Bob. Bob then performs a measurement on $\Psi$ with one randomly selected basis ($\sigma_x$ or $\sigma_z$). If Bob chooses the same basis as Alice, their results will correlate and thereby be the same. However, if Bob chooses a different basis from Alice, his result will be random.
	\item \textbf{Repetition and storing of states:} Alice and Bob repeat the prior step \textit{n} times to get the raw key. For every iteration, they store the state $\Psi$ and the basis $\sigma_{x/z}$ in a list for the raw key.
	\item \textbf{Deletion of corrupted states:} Bob informs Alice about which qubits he has recieved. Due to fiber attenuation, detector efficiency and other noise, many photons do not even reach their destination at Bob. This is why Bob needs to inform Alice so that she can delete all the lost states from her list.
	\item \textbf{Key sifting:} In this step Alice and Bob discuss which basis they have chosen for their measurements for every state via the classical channel. If their chosen basis is unequal, both delete the corresponding states from their list. Note that Alice and Bob do not share any measurement results with each other. Theoretically, Alice and Bob should now share the same sifted key. For BB84 the sifted key is $\frac{1}{2}$ of the size of the raw key.
	\item \textbf{Security check:} Unfortunately, due to noise and also due to a possible eaversdropper, the states Bob receives could differ from the ones Alice sent. This is why Alice and Bob compare some of their correct measurements and calculate the Quantum Bit Error Ratio (QBER). The calculation of QBER is introduced later. What is currently important to note is that the presence of Eve leads to a QBER of up to 25\%. In other words, QBER is an indicator as to whether the connection is secure. 
	\item \textbf{Error correction and privacy amplification:} Since we cannot distinguish between channel noise or the presence of Eve by the evaluation of QBER, error correction and privacy amplification measures are taken to reduce the information Eve $I_{ae}$ might have gained. After this last step, Alice and Bob share a secret key as $I_{ab}>I_{ae}$.
\end{enumerate}

Eve's increasing QBER stems from the fact that $\Psi$ changes its state when Eve measures in the wrong basis. However, if, in this case, Bob chooses the same basis as Alice, there is a 50\% chance of him receiving a false result. Since there is a 50\% chance that Eve will choose the wrong basis, there is a 25\% chance that Alice's and Bob's results will be diverse ($50\%*50\%=25\%$), which shows up in QBER \cite{bennett1984quantum}. The randomness of Alice's states is essential for security, which is why a true random generator is another requirement of a secure QKD protocol.\\
Since Eve can only measure once, the following question arises: Could Eve create multiple copies of the quantum state to measure multiple times? Eve could then select the right result once Alice has shared her MBs with Bob. Fortunately, this scenario is impossible due to the fundamental principles of quantum mechanics. The No-Cloning Theorem, which states that quantum information cannot be copied, precludes such an approach. This theorem has been widely recognized in QKD and has remained unchallenged since its introduction in 1982 \cite{wootters1982cloning}.

An important security point is that only one photon exists per state. Two or more photons carrying the same information would skew security, as Eve would be able to perform several measurements on the same $\Psi$. However, if Alice's photon source has imperfections and sometimes emits 2 photons, this could harm security. In that case, Eve could perform photon number splitting (PNS) \cite{scarani2004sarg04}.

\newpage
\subsubsection*{COW}

\begin{comment}
\begin{figure}[h]%[h!]
  \begin{center}
    \includegraphics[width=0.9\textwidth]{graphics/COW.png}
    \caption{COW Protocol (draw myself or ask  about permission)}
    \label{fig:cow}
  \end{center}
\end{figure}
\end{comment}

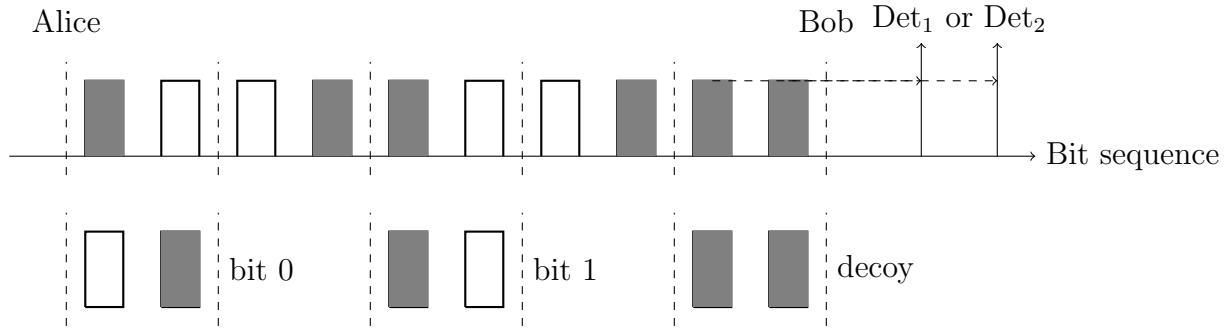
\begin{figure}
\begin{center}
\begin{tikzpicture}

% Time axis
\draw[->] (0, 0) -- (13.5, 0) node[right] {Bit sequence};

% Pulses for the COW protocol
\draw[thick] (1, 0) -- (1, 1) -- (1.5, 1) -- (1.5, 0);
\draw[thick] (2, 0) -- (2, 1) -- (2.5, 1) -- (2.5, 0);
\draw[thick] (3, 0) -- (3, 1) -- (3.5, 1) -- (3.5, 0);
\draw[thick] (4, 0) -- (4, 1) -- (4.5, 1) -- (4.5, 0);
\draw[thick] (5, 0) -- (5, 1) -- (5.5, 1) -- (5.5, 0);
\draw[thick] (6, 0) -- (6, 1) -- (6.5, 1) -- (6.5, 0);
\draw[thick] (7, 0) -- (7, 1) -- (7.5, 1) -- (7.5, 0);
\draw[thick] (8, 0) -- (8, 1) -- (8.5, 1) -- (8.5, 0);
\draw[thick] (9, 0) -- (9, 1) -- (9.5, 1) -- (9.5, 0);
\draw[thick] (10, 0) -- (10, 1) -- (10.5, 1) -- (10.5, 0);

% Decoy pulses (filled)
\fill[gray] (1, 0) rectangle (1.5, 1);
\fill[gray] (4, 0) rectangle (4.5, 1);
\fill[gray] (5, 0) rectangle (5.5, 1);
\fill[gray] (8, 0) rectangle (8.5, 1);
\fill[gray] (9, 0) rectangle (9.5, 1);
\fill[gray] (10, 0) rectangle (10.5, 1);

% Labels
\node[above] at (0.75, 1.5) {Alice};
\node[above] at (10.75, 1.5) {Bob};

% Vertical lines between every second pulse
\draw[dashed] (0.75, -0.25) -- (0.75, 1.25);
\draw[dashed] (2.75, -0.25) -- (2.75, 1.25);
\draw[dashed] (4.75, -0.25) -- (4.75, 1.25);
\draw[dashed] (6.75, -0.25) -- (6.75, 1.25);
\draw[dashed] (8.75, -0.25) -- (8.75, 1.25);
\draw[dashed] (10.75, -0.25) -- (10.75, 1.25);

% Detectors
\draw[->] (12, 0) -- (12, 1.5);
\node[above] at (12.5, 1.5) {Det\textsubscript{1} or Det\textsubscript{2}};
\draw[->] (13, 0) -- (13, 1.5);

% Pulses to detectors
%\draw[->, dashed] (1.25, 1) -- (11, 1);
%\draw[->, dashed] (2.25, 1) -- (12, 1);
%\draw[->, dashed] (4.25, 1) -- (11, 1);
\draw[->, dashed] (9.25, 1) -- (13, 1);
\draw[->, dashed] (10.25, 1) -- (12, 1);

%draw legend

\draw[dashed] (0.75, -2.25) -- (0.75, -0.75);
\draw[thick] (1, -2) -- (1, -1) -- (1.5, -1) -- (1.5, -2) -- (1, -2);
\draw[thick] (2, -2) -- (2, -1) -- (2.5, -1) -- (2.5, -2) -- (2, -2);
\fill[gray] (2, -2) rectangle (2.5, -1);
\draw[dashed] (2.75, -2.25) -- (2.75, -0.75);
\node[right] at (2.75, -1.5) {bit 0};

\draw[dashed] (4.75, -2.25) -- (4.75, -0.75);
\draw[thick] (5, -2) -- (5, -1) -- (5.5, -1) -- (5.5, -2) -- (5, -2);
\draw[thick] (6, -2) -- (6, -1) -- (6.5, -1) -- (6.5, -2) -- (6, -2);
\fill[gray] (5, -2) rectangle (5.5, -1);
\draw[dashed] (6.75, -2.25) -- (6.75, -0.75);
\node[right] at (6.75, -1.5) {bit 1};

\draw[dashed] (8.75, -2.25) -- (8.75, -0.75);
\draw[thick] (9, -2) -- (9, -1) -- (9.5, -1) -- (9.5, -2) -- (9, -2);
\draw[thick] (10, -2) -- (10, -1) -- (10.5, -1) -- (10.5, -2) -- (10, -2);
\fill[gray] (9, -2) rectangle (9.5, -1);
\fill[gray] (10, -2) rectangle (10.5, -1);
\draw[dashed] (10.75, -2.25) -- (10.75, -0.75);
\node[right] at (10.75, -1.5) {decoy};

\end{tikzpicture}
\caption{Illustraition of COW Protocol}
\label{fig:cow}	
\end{center}
\end{figure}

The Coherent One-Way (COW) protocol is a QKD protocol designed to be simpler and more robust against practical imperfections compared to BB84. Given its extensive use in our work, we discuss its steps in detail here \cite{gisin2004cow}:

\begin{enumerate}
    \item \textbf{Preparation}: Alice prepares ``bit 0" or ``bit 1" with probability $(1-f)/2$ and decoy sequences with probability $f$. Alice uses a mode-locked laser emitting pulses with a mean photon number $\mu$, spaced by a fixed time interval $\tau$, ensuring phase coherence between non-empty pulses. A variable attenuator selectively blocks some pulses, potentially reducing the setup to a continuous-wave laser followed by an attenuator. Each logical bit is encoded in a two-pulse sequence as illustrated in Figure~\ref{fig:cow}. This encoding means that a sequence of pulses represents a string of bits.
    \item \textbf{Exchange}: The encoded pulses are transmitted to Bob via a quantum channel with a transmission coefficient $t = 10^{-\alpha d/10}$ (where $\alpha$ is the attenuation coefficient in dB/km and $d$ is the distance). Bob's setup includes a non-equilibrated beam-splitter with a transmission coefficient $t_B$. The transmitted pulses are used to establish the raw key. Bob needs to distinguish between the two non-orthogonal states \cite{gisin2004cow}, which are represented by logical bits. Unambiguous discrimination between these states is achieved by photon counting, with a success probability of $p_{ok} = 1 - e^{-\mu t t_B}$ (with $\alpha = \sqrt{\mu t t_B}$).
    \item \textbf{Public Discussion}: Alice reveals the decoy sequences. Bob removes corresponding detections from his raw key and checks for monitoring line detections to estimate coherence. Pulses reflected by the beam-splitter go to an interferometer to detect eavesdropping by measuring coherence. In the presence of coherence, detector 2 should not fire for certain pulse pairs, indicating no eavesdropping.
    \item \textbf{Error Correction and Privacy Amplification}: To counter specific eavesdropping strategies, Alice introduces decoy sequences by leaving both pulses in a pair non-empty with a probability $f$. These sequences do not encode bit values but serve to check for eavesdropping. Decoy pulses have a different $\mu$ that regular pulses. However, Eve cannot distinguish between them. The sequence is discarded if a decoy sequence is detected on the data line. Detection in the monitoring line ensures coherence is maintained. Eve’s attacks would otherwise disturb this coherence, revealing her presence.
\end{enumerate}

The protocol accounts for errors due to dark counts and reduced visibility in the monitoring interferometer. Imperfections in detectors and other components are considered to ensure realistic security analysis. Strategies such as intercept-resend (I-R) and photon-number-counting attacks (PNC) are analyzed, and decoy sequences help detect these attacks \cite{gisin2004cow}.

The protocol's simplicity means fewer components and potentially lower costs. The use of standard telecom devices makes it accessible for practical implementations. The monitoring line effectively detects eavesdropping, ensuring robust security \cite{gisin2004cow}.
The proposed protocol strikes a balance between theoretical rigor and practical feasibility, addressing key challenges in the implementation of secure QKD systems.

\newpage
\section{State-of-the-art Solutions}

QKD offers secure communications, even against future quantum computers, and has evolved from single links to extensive networks. Recent developments in Europe focus on real-world use cases and collaboration with end-users to demonstrate QKD's practical deployment and highlight its security advantages \cite{hubel2023deployed}.
Selecting a QKD system necessitates the careful choice of an appropriate protocol. In the following section, we will elaborate on the various protocol options available.

The BB84 protocol \cite{bennett1984quantum} puts high specifications onto the QKD hardware. Single photon sources required to ensure security discussed in \cite{shor2000securitybb84} have very complex setups (Ion trap \cite{keller2004iontrap}) and high operating costs (superconducting single-photon detectors (SSPDs) \cite{goltsman2001superconducting}, also refered to as superconducting nanowire single-photon detectors (SNSPDs) \cite{natarajan2012superconducting}, require a temperature of 4K to operate). A simpler alternative is weak laser pulses. If a laser is operated with very low powers, it technically only emits one photon at a time. The only problem here is that the photons per pulse are Poisson distributed \cite{shor1983subpoissonian}\cite{lounis2000single}\cite{lounis2005single}. Under these circumstances the security of BB84 cannot be guaranteed \cite{shor2000securitybb84}. This raises the need for another protocol to handle these imperfections better than BB84. 
SARG04 is one such protocol that is robust against PNS attacks \cite{scarani2004sarg04}. However, the disadvantage of SARG04 is that only $\frac{1}{4}$ of the raw keys remain after the sifting procedure.
Other well-known QKD protocols were discussed in \cite{nurhadi18qkdsurvey}, such as BB84, E91, BBM92, B92, SARG04 and COW to name a few. Some of these protocols are prepare-and-measure (e.g. BB84, B92) based, while others are entanglement-based (e.g. BBM92). Additionally, the authors conducted simulations on BB84, B92 and BBM92 to analyze the errors per key for each protocol.

The COW protocol is a QKD protocol designed to be simpler and more robust against practical imperfections compared to other protocols, such as BB84. It was introduced by Nicolas Gisin and his collaborators in 2004 \cite{gisin2004cow}. The COW protocol is particularly notable for its suitability in real-world implementations due to its straightforward detection scheme and resilience to PNS attacks. The COW protocol utilizes coherent states of light for key distribution which are more easily generated and detected with current technology compared to single photons. As shown in Figure~\ref{fig:cow}, the protocol encodes information in the presence or absence of coherent pulses. Shor claims in \cite{shor2000securitybb84} that weak-coherent protocols had not been proven to be secure. However, Lo \cite{lo2005decoy} demonstrated that decoy state QKD combines the benefits of both worlds: unconditional security and high performance.
%This claim is supported and extended to a more generalized szenario by Reutov \cite{reutov2023securityDecoy} under the condition that decoy-state pulses are used by the QKD protocol.

In this chapter, we investigated various QKD protocols. The protocols employing the COW, which also concern different equations for SKR and are particularly relevant to our research, include \cite{stucki2009long}, \cite{eraerds2010quantum} and \cite{walenta2014fast}. Despite the similarities in setup and fundamental principles, we observed distinct equations for the SKR due to different assumptions and focuses on different parameter impacts. Moving forward, we will critically compare the solutions proposed in these papers, selecting and adapting the most suitable model for our implementation. This approach ensures that our model not only aligns with our specific requirements but also incorporates the most effective strategies for the analysis of the SKR.

%\cite{stucki2005fast}

\newpage
\section{Objectives}

%\subsection{Purpose Statement}

\begin{figure}[h]%[h!]
  \begin{center}
    \includegraphics[width=0.49\textwidth]{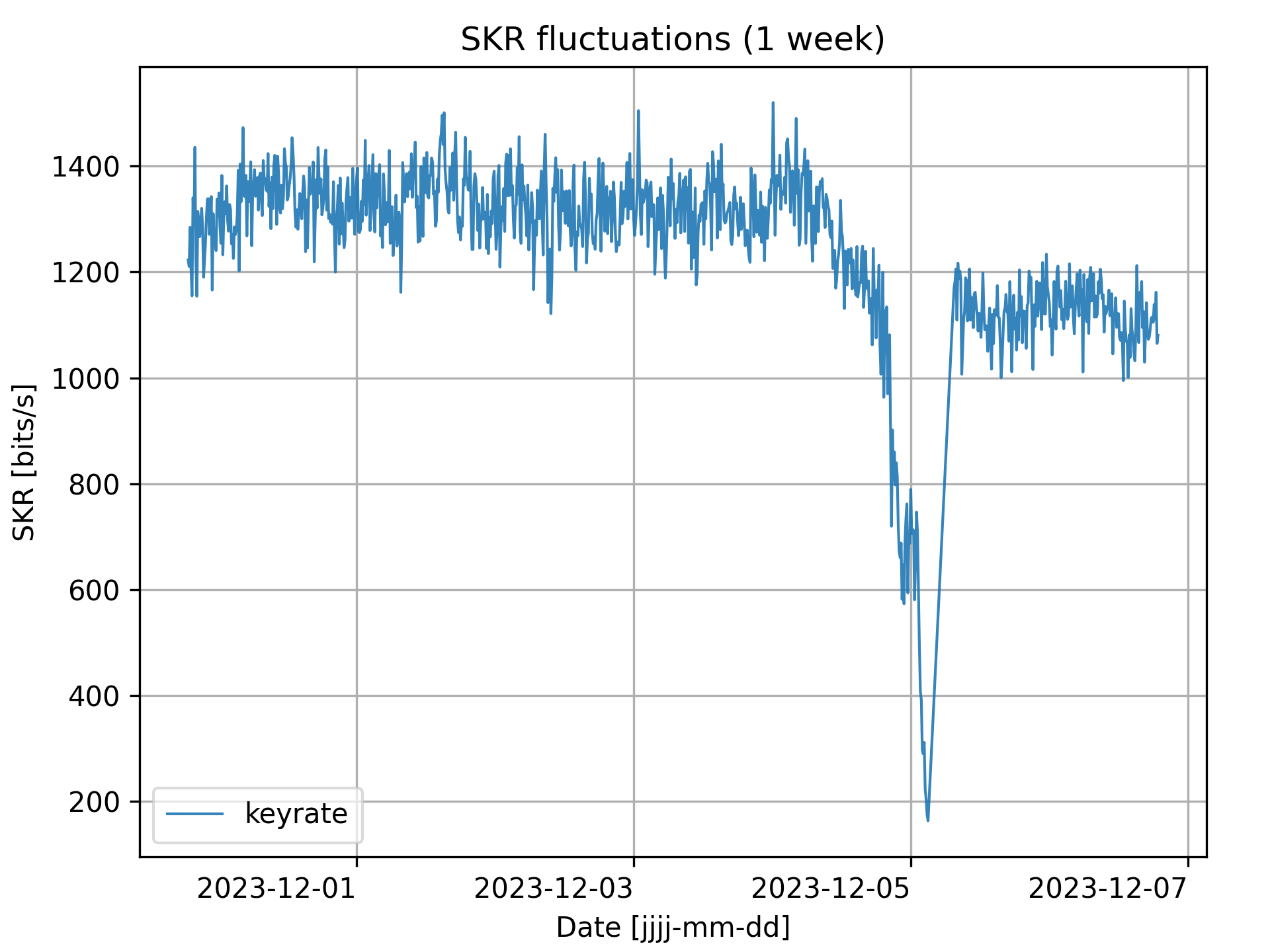}
    \includegraphics[width=0.49\textwidth]{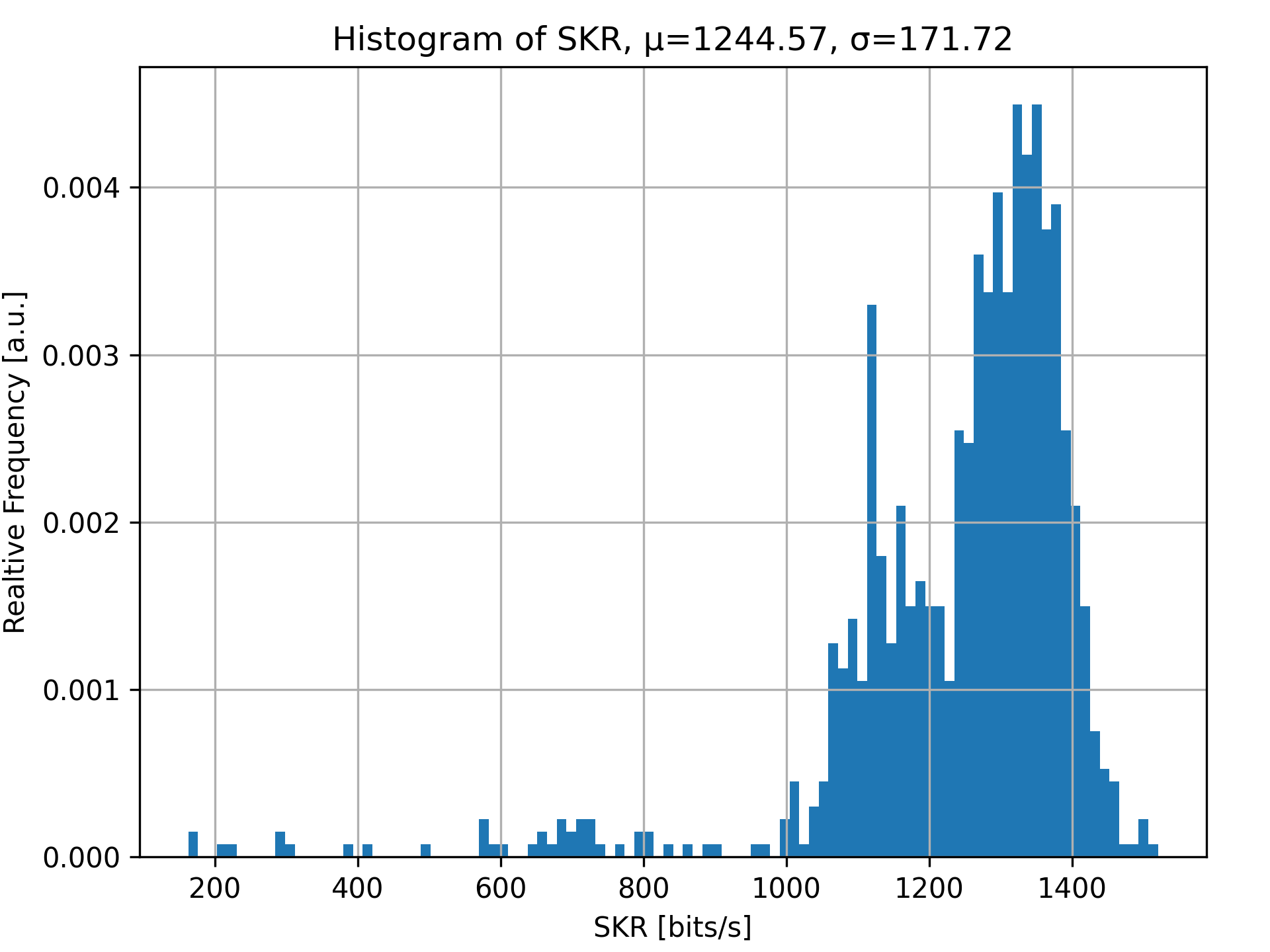}
    \caption{Measured SKR of link 1 with respect to time and histogram}
    \label{fig:SKRhistogram}
  \end{center}
\end{figure}

Despite the advantages of the COW protocol in practical implementations of QKD, it faces several significant challenges due to system imperfections. These challenges include low SKR, fluctuating and variable SKR, and unstable connections. To date, there is a noticeable gap in the research addressing, analyzing, or providing solutions for these specific issues. In our study, we have observed unstable SKR behaviors in QKD systems, as illustrated in Figure~\ref{fig:SKRhistogram}. Given the lack of sophisticated studies on this behavior, we have undertaken the task of addressing these problems ourselves.

Figure~\ref{fig:SKRhistogram} consists of two subplots: The left subplot illustrates the measured Secret Key Rate (SKR) over time, while the right subplot shows the statistical distribution (histogram) of the SKR. Both subplots represent the same QKD link and reveal a high variability of the SKR over time. The higher the SKR, the better.

This variability in SKR raises several important research questions that have not yet been answered: First, what are the key practical imperfections in QKD systems and how do they deviate from idealized protocol conditions, potentially compromising security and efficiency? Understanding these imperfections is crucial for improving QKD system performance.
Secondly, how can monitoring data be effectively utilized to develop a physical layer model for QKD that accurately reflects the influence of these imperfections? By leveraging monitoring data, we aim to create a model that can account for the real-world conditions affecting QKD systems.\\
Finally, to what extent can the developed physical layer model predict QKD system performance and security outcomes under varying conditions? We aim to determine whether we can anticipate the observed fluctuations in SKR and identify the underlying reasons for these variations.
By addressing these questions, we aim to enhance the reliability and security of QKD systems, ensuring their robustness in real-world applications.

\chapter{Methodology}

Chapter 3 summarizes the methodologies used in this thesis. First, the measurement setup and data collection are described. Second, a model of the physical layer is developed, and third, the implementation process of these models is discussed.

\section{Methodology for Data Collection} 

This section discusses our measurement setup and how the data is collected. Further, it describes all relevant parameters and how these parameters are preprocessed.

\subsection{Monitoring Equipment and Data Acquisition} 

This section focuses on the QKD network we are working with and on data collection process.

\subsubsection{Monitoring Equipment}

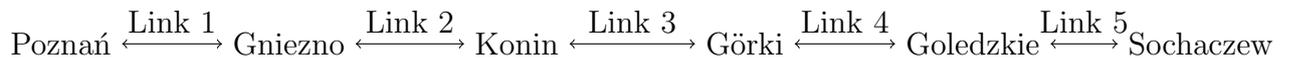
\begin{figure}[h]%[h!]
	\begin{center}
		\begin{tikzpicture}
    		% Nodes
    		\node (Poznan) at (0,0) {Poznań};
    		\node (Gniezno) at (3,0) {Gniezno};
    		\node (Konin) at (6,0) {Konin};
    		\node (Gorki) at (9,0) {Görki};
    		\node (Goledzkie) at (12,0) {Goledzkie};
    		\node (Sochaczew) at (15,0) {Sochaczew};
    		%\node (Warszawa) at (18,0) {Warszawa};
    		% Edges
    		\draw[<->] (Poznan) -- (Gniezno) node[midway, above] {Link 1};
    		\draw[<->] (Gniezno) -- (Konin) node[midway, above] {Link 2};
    		\draw[<->] (Konin) -- (Gorki) node[midway, above] {Link 3};
    		\draw[<->] (Gorki) -- (Goledzkie) node[midway, above] {Link 4};
    		\draw[<->] (Goledzkie) -- (Sochaczew) node[midway, above] {Link 5};
    		%\draw[<->] (Sochaczew) -- (Warszawa) node[midway, above] {Link 6};
			\end{tikzpicture}
    	\caption{QKD Network Setup in Poland}
    	\label{fig:setuppoland}
	\end{center}
\end{figure}

The analyzed data comes from a deployed QKD network in Poland mentioned in \cite{hubel2023deployed}. The evaluated network extends from Poznan to Sochaczew. Figure~\ref{fig:setuppoland} shows the connections between the cities and elucidates the links.  
%Poznan – Gniezno – Konin – Görki – Goledzkie – Sochaczew – Warszawa  \\

\begin{table}%[h]
	\centering
	\begin{tabular}{|c|c|c|c|c|}
		\hline
		Link 1 & Link 2 & Link 3 & Link 4 & Link 5 \\		
		\hline
		%Poznan-Gniezno & Gniezno-Konin & Konin-Görki & Görki-Goledzkie & Goledzkie-Sochaczew\\
 		%\hline
		46 km  & 57 km & 42 km & 43 km & 50 km \\
		\hline
	\end{tabular}
	\caption{Air-line distances for all links}
	\label{tab:distances}
\end{table}

The system does not make use of entanglement swapping and instead uses a prepare-and-measure QKD protocol for each link \cite{gisin2004cow}. Therefore, each link's end must be a trusted node\cite{wenning2023towards}. Each node first decrypts the end-to-end key with one key data as a receiver (Bob) and then encrypts the end-to-end again with a different key as the sender (Alice) \cite{johann2024comparison}.\\
Every link uses a separate dark fiber. We define dark fiber as there being no other data traffic on the fiber except for QKD key exchanges. All the links exchange keys and monitor several parameters continuously.

The Cerberis\textsuperscript{3} QKD system from ID Quantique is used as hardware \cite{idq2019cerberis3}. This system uses the COW protocol with a pulse repetition rate of 1.25GHz. The Cerberis\textsuperscript{3} typically has an SKR of 1400 bits/s @12 dB.
The maximum length of the quantum channel (typ. @ 0.23 dB/km) is given with 50 km (@ 12 dB, optionally they offer device up tp 70/80 km or @ 16/18 dB on request) \cite{idq2019cerberis3}.

\subsubsection{Data Acquisition}
This chapter establishes a secure data acquisition process for monitoring the QKD system. To ensure secure data transmission, a VPN tunnel is configured to the Poznan Supercomputing and Networking Center (PSNC). The Simple Network Management Protocol (SNMP) is utilized within this infrastructure to facilitate the identification of specific parameters critical to assessing the performance and security of the QKD system. These parameters are defined by unique identifiers (IDs) in the SNMP setup. InfluxDB is the primary data infrastructure for effective data management and storage. This database is suitable for time-series data, offering a solution for systematically capturing and analyzing the QKD system's operational data.\\

The data is retrieved from the server after a monitoring period longer than one week has passe. All files are CSV documents and each parameter is stored in a separate file. The CSV file contains a chronological list of the measured parameter (e.g., key rate), with each value assigned a date and time stamp. For example, a line in the key rate file has the following format: 2023-11-29 18:57:02.113707+00:00,1197\\
These files are stored locally for further processing. The data processing is discussed in ~Section~\ref{sec:dataprep}.\\

\subsection{Parameters Monitored}% and Data Preprocessing Techniques} 
\label{sec:params}
%\subsubsection{Parameters}

For the considered QKD network, we assess five parameters limited by the monitoring capabilities of the QKD devices: Secret Key Rate (SKR), Quantum Bit Error Rate (QBER), visibility and laser power. These metrics are fundamental to both the operational efficiency and the security posture of the QKD system.

\subsubsection*{SKR}
The SKR, measured in bits per second, serves as a primary indicator of the system's performance in secret key transmissions per second between two nodes, Alice and Bob. It quantifies the volume of quantum keys exchanged over a given period, directly influencing the rapidity and robustness of secure communications. Optimizing the key rate is crucial to maximizing secure data throughput \cite{eraerds2010quantum} since, in OTP, the keys have to match the size of the information, which means that the SKR of the QKD limits the performance of the OTP protocol \cite{shannon1949communication}.

\subsubsection*{QBER}
QBER is defined as the ratio of erroneously transmitted qubits divided by the total transmitted qubits. It serves as a measure of transmission accuracy (see ~Equation~\ref{eq:QBER}) and system integrity.  Lower QBERs indicate higher transmission fidelity. Given its inverse relationship with the key rate, the QBER is a crucial metric for assessing the efficiency and security of the QKD process, as a rising QBER indicates a higher rate of incorrect detected information on Bob's side. Incorrect detections can stem either from a non-stable channel or from Eve trying to eavesdrop. Unfortunately, it is not possible to distinguished between a system's fault or a fault caused by Eve, which is why methods like privacy amplification exist to increase security after the key exchange. Efforts to minimize the QBER are essential for improving the SKR. However, the reduction of the QBER is limited to adaptable system parameters \cite{eraerds2010quantum}:
\begin{equation}
    \begin{split}
        \label{eq:QBER}
        \textrm{QBER} &= \frac{\textrm{\# false detections}}{\textrm{\# total detections}}\\
    \end{split}
\end{equation}

\subsubsection*{Visibility}
In the context of QKD, visibility refers to the qualitative measure of the interference pattern. High visibility signifies well-differentiated quantum states, resulting in a higher SKR \cite{walenta2014fast}. According to \cite{walenta2014fast}, the visibility is calculated as follows:
\begin{equation}
    \begin{split}
        \label{eq:Visibility}
        V &= 1 - \frac{N_{\text{int}}}{N_{\text{non}}}  \frac{p_{\text{non}}}{p_{\text{int}}}
    \end{split}
\end{equation}
$N_{\text{int}}$ is the number of detections from sequences that should destructively interfere, and $N_{\text{non}}$ is the number of detections from non-interfering sequences. The $p_{\text{non}}/p_{\text{int}}$ is the ratio of non-interfering to interfering sequences sent.

\subsubsection*{Laser Power}
Laser power is closely linked to the emission rate of photons per pulse (introduced in Section~\ref{sec:cowparams}) which has an impact on the signal intensity and, thus, the SKR's and the protocol's security. Because the datasheet of the Cerberis\textsuperscript{3} does not provide any information about the unit measuring laser power, we assume that laser power is measuring photons per pulse. Measurements verify the data range of the average photon number per pulse. An increase in laser power can elevate the probability of photon detection, potentially enhancing SKR. However, it requires careful tuning to ensure that an increase in photon emission does not adversely provide more corroding surfaces for an eavesdropper to attack by PNS \cite{brassard2000limitations}.
%\textbf{Compression Ratio}\\
%The compression ratio within a QKD framework addresses the efficiency of quantum key compression and aims to maximize bandwidth utilization without compromising the keys' security attributes. An optimal compression ratio signifies an efficient key utilization strategy, facilitating the secure transmission of a greater volume of data. This efficiency also contributes to the communication channel's throughput \cite{walenta2014fast}.\\
The systematic evaluation and tuning of these parameters is crucial for developing secure, efficient, and scalable QKD systems.

\subsection{Data Preprocessing Techniques}
\label{sec:dataprep}

It is crucial to focus on specific techniques that enhance the quality and reliability of the data before model fitting. 
Therefore, this section will describe what methodologies are taken into account for preprocessing the data accordingly.

\subsubsection*{Data Importing}
The data is imported from CSV files into a Pandas DataFrame \cite{reback2020pandas} \cite{mckinney2010data}. This DataFrame is used in between every preprocessing step to temporarily store data.

\subsubsection*{Data Cleaning and Filtering}
NaN values are undesirable since they can disrupt calculations and can lead to unexpected behavior in functions.
To prevent this, all NaN values in the DataFrame are filtered and removed.\\
We used methods to detect and potentially remove outlier data points such as visibility data greater than 1 which is, by definition, a nonvalid value for visibility~\cite{walenta2014fast}. These outliers may arise from quantum or classical noise that could significantly skew the data analysis \cite{han22dataMining}.

\subsubsection*{Temporal Averaging}
Since the data is collected in a time-series format, averaging over time is suitable to reduce noise and stabilize the measurements. 
The appropriate window sizes are selected based on the stability and dynamics of the QKD system and the specific objectives for which the data is used, such as trend prediction. \cite{cleveland1988locallyRegression}\\
In order to minimize noise and focus on signal trends, a rather large window size of two to ten minutes is taken for averaging.

\newpage
\section{Development of the Physical Layer Model} 

Within this section we describe the theoretical model behind a QKD system based on the COW protocol.

%\subsection{Assumptions and Theoretical Justifications} %Rationale Behind the Model}

\subsection{Parameters Modeling the SKR}
\label{sec:cowparams}

In our study of QKD, several key parameters significantly influence the performance and security of the system. These Parameters are discussed in this section.

First, we investigate the fiber attenuation ($\alpha$), expressed in decibels per kilometer (dB/km), which quantifies the loss of signal strength as light travels through the fiber. A typical value for $\alpha$ is 0.21 dB/km \cite{eraerds2010quantum}.
The avalanche photodiode quantum efficiency ($\eta$) measures the effectiveness of the photon detectors in converting incident photons into electrical signals. A typical photodiode efficiency according to\cite{eraerds2010quantum} is $\eta=0.07$.\\
Error correction efficiency ($\eta_{ec}$), also expressed in percentage, is crucial in QKD post-processing to ensure secure communication.
Dark count (\(p_{dc}\)) refers to the false detection of photons by a photodiode due to thermal noise or electronic defects. This means that a photon is detected when there is none.
Crosstalk and Raman scattering are phenomena that can significantly impact the performance and security of QKD systems, only in non-dark fibers where classical and quantum communications coexist. For the sake of completeness, we discuss the influence of both QKD systems. Crosstalk refers to the unwanted transfer of signals between communication channels, leading to interference and degradation of the quantum signal. Concurrently, Raman scattering introduces noise by energy changes through photon scattering. Both phenomena can introduce errors and compromise the fidelity and security of the transmitted quantum keys, necessitating careful management and mitigation strategies in the design and operation of QKD systems. Both the Raman-scattering and the Crosstalk can be omitted in our model since our QKD network in Poland operates on dark fiber, meaning that there is no light traffic on the fiber other than the QKD laser pulses that are sent.
Storage length ($L_s$) in meters (m) indicates the maximum distance for storing quantum states within the communication setup. Typical values for this and the following parameters are mentioned in the next section in Table~\ref{tab:parametervalues}.
The loss in Bob's components ($t_B$) is measured in decibels (dB) to account for internal losses in Bob's receiving end of the QKD system.
The detector dead time ($\tau_{dead}$) in seconds (s) is the period during which the detector is unable to register further photon arrivals after detecting an event, impacting system throughput.
The repetition rate ($f_{rep}$ or $\nu$), expressed in Hertz (Hz), denotes the frequency at which quantum states are prepared and measured, directly influencing the system's data rate.
The average photon number per pulse ($\mu$) is a critical parameter balancing between security and system performance. It statistically represents the intensity of the light pulses, indicating the mean number of photons emitted in each pulse. For instance, a $\mu=0.5$ would mean that statistically, there is only one photon every two pulses, assuming our detector emits either 0 or 1 photon.
Lastly, we introduce the total fiber length ($L$), measured in kilometers (km), as a fundamental physical parameter that influences both the attenuation and the overall feasibility of the quantum communication link.

\subsection{Ranges of Parameters}

Table \ref{tab:parametervalues} states all parameters with their corresponding value and value range. We chose the value ranges according to the measured data, typical values from \cite{eraerds2010quantum}, and theoretical bounds.

\begin{table}[h]
	\centering
	\begin{tabular}{|c|c|c|c|}
		\hline
		Parameter & Symbol & Value & Test Range  \\
		\hline
		Fiber attenuation & $\alpha$ & 0.21 dB/km & [0.15, 0.25]dB/km \\
        \hline
        Detection Efficiency & $\eta$ & 0.07 & [0.02, 0.1] \\
        \hline
        Error Correction Efficiency & $\eta_{\text{ec}}$ & 1 & [1, 6/5] \\  
        \hline   
        Dark Count Probability & $p_{\text{dc}}$ & $5*10^{-6}$ & [$2.5*10^{-6}, 10*10^{-6}$] \\  
        \hline
        Storage Length & $L_{\text{s}}$ & 10 km & 10 km \\ 
        \hline
        Component Loss (Bob's) & $t_{\text{B}}$ & $10^{-2.65/10}$ & $[0.5 * 10^{-2.65/10},2 * 10^{-2.65/10}]$ \\ 
        \hline
        Detector Dead Time & $\tau_{\text{dead}}$ & $10*10^{-6}$ s & $[1*10^{-6}, 10*10^{-6}]$ s \\ 
        \hline
        Channel Length & $L$ & 40 km & up to 100km \\
        \hline 
        Pulse Repetition Rate & $f_{\text{rep}}$ & $1.25*10^{9}$ Hz & - \\ 
        \hline
        Mean Photon Number & $\mu$ & 0.5 & [0.01, 0.5] \\ 
        \hline
	\end{tabular}
	\caption{Parameters and their assigned values}
	\label{tab:parametervalues}
\end{table}

\subsection{Equations Based on the COW Protocol}

In the following, we derive the equations for the SKR. The equations are based on \cite{eraerds2010quantum} and are simplified with the consideration of having a dark fiber for QKD.\\

Auxiliary Equations:

\begin{equation}
    \begin{split}
        \label{eq:COWtpd}
        t &= 10^{-\alpha L/10}\\
        P &= \frac{1}{2} + \sqrt{D \ (1 - D)}\\
        D &= \frac{1 - V}{2 - \frac{\mu}{t}} \\
    \end{split}
\end{equation}

Detection rate after Key Sifting:
\begin{equation}
    \begin{split}
        \label{eq:COWrsift}
        R_{Sift} &= \frac{1}{2} (\beta \  p_{\mu} + 2 \ p_{dc} + p_{AP}) \ \nu \ \eta_{duty} \ \eta_{dead}\\
    \end{split}
\end{equation}
Compared to the theoretical model of \cite{eraerds2010quantum} some simplifications were made in this equation regarding the Raman scattering and the Cross-Talk as our used hardware operates on dark fiber.\\

Shared information between Alice and Bob:
\begin{equation}
    \begin{split}
        \label{eq:COWiab}
        I_{ab} &= 1 - \eta_{ec} \ H_2(QBER) \\
    \end{split}
\end{equation}

Shared information between Alice and Eve:
\begin{equation}
    \begin{split}
        \label{eq:COWiae}
        I_{ae} &=  \frac{(1 - \frac{\mu}{2t}) \ (1 - H_2(P)) + \frac{\mu}{2t}}{1 + \frac{2 \ p_dc}{(\mu \ t \ \eta)}} \\
    \end{split}
\end{equation}

Secret Key Rate:
\begin{equation}
    \begin{split}
        \label{eq:COWskr1}
        SKR &=  R_{Sift}\ (I_{ab}-I_{ae}) \\
    \end{split}
\end{equation}

%\subsubsection{Simplified COW Equations Assuming BB84 Coding}
\subsection{Assumptions and Theoretical Justifications for Simplified Protocol}
\label{Met:COWSimplified}

Since we aim to use a simplified protocol which is easier to implement, we set $\beta=1$ and approach the upper bound by setting $\mu = t$. \cite{eraerds2010quantum}
We approach the equation of the SKR by the following general equation: $SKR = r \ (b - e)$.

Introducing $\eta_{dead}$ and  $\eta_{duty}$ as
\begin{equation}
    \begin{split}
        \label{eq:COWeta}
        \eta_{dead} &= \frac{1}{1 + (p_{\mu} + 2 p_{dc} + p_{AP}) \ \nu \ \tau_{dead}}\\
        \eta_{duty} &= \frac{L_s}{L + 2* L_s}
    \end{split}
\end{equation}
and applying it onto equation \ref{eq:COWrsift} and \ref{eq:COWskr1} leads to our final equation for the SKR.
\begin{equation}
    \begin{split}
        \label{eq:COWskr}
        SKR = &\frac{\nu (p_{\mu} + 2 \ p_{dc} + p_{AP}) L_s}{2 (L+2L_s)(1+\tau_{dead} \nu (p_{\mu} + 2 \ p_{dc} + p_{AP}))} \\ 
        &\lbrack 1 - \eta_{ec} H(\frac{(1-V) p_{\mu} + 2 \ p_{dc} + p_{AP}}{2 (p_{\mu} + 2 \ p_{dc} + p_{AP})}) - \frac{(1-\frac{\mu}{2t})(1-H(P(V,\mu,t))+\frac{\mu}{2t}}{1+\frac{2P_{dc}}{\mu t \eta}}  \rbrack \\
    \end{split}
\end{equation}

\subsection{QKD Protocols and Perfomances}

%doi:10.48550/arXiv.quant-ph/050609 ?

Since the specific protocol of our use QKD device is not explicitly stated in the provided data sheet \cite{idq2019cerberis3}, but other devices from IDQ employ the COW protocol \cite{idq2024cerberisXG}, we examined different QKD protocols described in the literature, including those by Stucki et al. \cite{stucki2009long}, Eraerds et al. \cite{eraerds2010quantum}, and Walenta et al. \cite{walenta2014fast}.\\
Despite similarities in setup and fundamental principles, each protocol presents different equations for calculating the SKR. By applying the equations from these sources \cite{stucki2009long}, \cite{eraerds2010quantum}, \cite{walenta2014fast} and using the values from the data sheet \cite{idq2019cerberis3}.\\
We determined that the best fit for our data aligns with the protocol described by Eraerds et al. \cite{eraerds2010quantum}.
Consequently, our focus has been on this particular implementation, which we have detailed extensively in previous chapters. This approach ensures a thorough understanding and accurate representation of the QKD system's performance under the assumed protocol.

\subsection{Approach for Evaluating the Distribution of $\mu$}

The SKR, as shown in Equation~\ref{eq:COWskr}, is a complex equation allowing little room for simplifications, as discussed in \cite{eraerds2010quantum}. Since it is too complex to simplify and solve analytically while preserving mathematical correctness, an automated approach to solve the SKR for $\mu$ is used instead. $\mu$ is calculated with a numeric solver function by setting all other parameters to a default value and then solving the equation \ref{eq:COWskr} for $\mu$.

\section{Model Implementation}

\subsection{Tools and Software Used}

%vs code\\
%anaconda\\
%python: pandas, sci-kit, scipy,\\
%tensorflow (+ cuda, tensorRT): keras\\

In the course of this research a variety of tools and software were utilized to manage, process and analyze the data, as well as to develop and validate the theoretical models. Below is a description of each tool and software package used:\\

%\subsubsection*{Visual Studio Code (VS Code)}
%Visual Studio Code was employed as the primary code editor for this project. Its extensive library of extensions and support for Python made it an ideal choice for writing, debugging, and managing the project's codebase. VS Code's integration capabilities with Git also facilitated version control, allowing for efficient management of changes and collaboration.

%\subsubsection*{Anaconda}
%Anaconda, a Python distribution, is used for managing the project's dependencies and environments. It simplified the process of setting up multiple isolated environments for different aspects of the project, ensuring that dependency conflicts were minimized. Anaconda's package manager, Conda, is helpfull in installing, running, and updating Python and the libraries needed for the research.

\subsubsection*{Python Libraries Used}
Python is the primary programming language used in this thesis due to its widespread adoption in scientific computing and its extensive ecosystem of libraries tailored for data analysis and machine learning while also being open source:
\begin{itemize}
    \item \textbf{Pandas}: This library is utilized for data manipulation and analysis. It provides powerful data structures to work with time-series data which is crucial for handling the outputs from QKD systems.
    \item \textbf{SciPy}: Used for scientific and technical computing, particularly because of its optimization and linear algebra (curve fitting) \cite{scipy2020}.
    \item \textbf{Scikit-learn}: This machine learning library is used for the data processing (e.g. Min-Max-Scaler, Grid search \cite{scikit-learn}) to prepare the data accordingly before implementing and evaluating the machine learning algorithms detailed in the methodology.
\end{itemize}

\subsubsection*{TensorFlow: Keras}
To build and train deep learning models, we use TensorFlow \cite{tensorflow2015-whitepaper} \cite{abadi2016tensorflow} supported by CUDA \cite{cuda} and TensorRT for enhanced performance on NVIDIA GPUs. This setup allowed for the leveraging of computational graphs, extensive parallel computing, and efficient training of models on large datasets.

As a high-level neural networks API that is running on top of TensorFlow, \textbf{Keras} \cite{keras} simplified the ML model's construction, training and evaluation. Its user-friendly interface accelerated the development process and enhanced model training reproducibility.

These tools and software packages provided a framework for addressing the data analysis and model development challenges encountered in this thesis on QKD.

%\subsubsection*{Documentation and Version Control}
%To keep track of all code changes git was used.

\subsubsection*{Technical Challanges}

In addressing the challenge of reproducibility in our machine learning models used for analyzing QKD systems, we identified variability in outcomes with each program execution as a primary concern. To mitigate the randomness of results, we initially set random seeds in our computational environment. This approach however, while effective in reducing variability, did not fully address the inherent non-determinism introduced by GPU computing, where not all algorithms are guaranteed to operate deterministically. Using the GPU with CUDA significantly enhances the model computations, whereas deactivating GPU computation leads to vital performance reductions. To overcome the challenge of receiving reproducible results, we utilized an experimental feature in Keras that allowed for the deactivation of non-deterministic functions on the GPU, thus securing the reproducibility of results on any single machine. However, this solution is not without limitations, as results still vary across different machines and hardware configurations specifically when code is executed on CPUs or GPUs.\\
Another challenge arose when trying to parallelize the training of different links with a driver that executes multiple ML models simultaneously. In this scenario, the GPU is likely to run out of memory as every model tends to allocate a substantial amount of memory at the program's start. This behavior could be addressed with a setting that limits memory allocation at startup and allows for dynamic memory allocation as needed during runtime.

\subsection{Technical Details on the Implementation Process}
\subsubsection*{Customized Tools and Libraries}
To better understand the collected data, a customized COW class reuses and adapts the underlying model. 
The main target of this class is to implement all mentioned system parameters of \cite{eraerds2010quantum} in a theoretical model and calculate the resulting SKR depending on QBER and visibility. The model results are shown in the Results chapter.

\subsubsection*{Theoretical Model Fitting}
The specific characteristics of quantum communications are considered so as to select the appropriate theoretical models that will fit the preprocessed data \cite{bishop2006pattern}. In this, theoretical relations between parameters are examined where we try to map the theoretical model onto the measured data.\\
The theoretical model that now accurately represents the underlying physical processes is then used to estimate parameters from the cleaned and averaged data.\\
The fitting of the parameters is done with the curve fit function from SciPy \cite{scipy2020}.

\subsubsection*{Data Preparation for ML}
To prepare the input data for the ML model, we used the Sci-kit Learn MinMaxScaler \cite{scikit-learn} to scale the SKR and the other parameters into a range between 0 and 1. This scaling process is essential because it standardizes the input data, ensuring that all features contribute equally to the learning process. We observed that when we scaled the data accordingly, the training error decreased significantly, demonstrating the importance of proper data normalization in improving the ML model's performance and convergence speed.

\subsubsection*{Model Architecture and Implementation Details}
In this study, a machine learning model utilizing a multi-input architecture is designed to handle the complexities inherent in quantum communication metrics.

The model itself is structured around four distinct input layers, each crafted to process a specific type of data input. Input layers \texttt{InputA}, \texttt{InputB} and \texttt{InputC} handle scalar values representing QBER, visibility and the link loss, while \texttt{InputD} processes vector-based features such as historical data arrays. This diverse input strategy allows for a comprehensive analysis of both current and historical QKD system performance metrics.
The link loss parameter is introduced to capture the loss of the network components within one parameter that varies between links. This helps to distinguish between data from different links, which enhances the accuracy and generality of the model training.

Each input stream passes through two Dense layers configured with Rectified Linear Unit (`ReLU') activation functions to introduce non-linearity and enable the model to capture complex data patterns. The layers progressively reduce in size from 64 neurons down to 16, which helps in refining the data to its most essential features before integration. This architectural choice reflects a strategic layering intended to maximize learning efficiency and data representation.

After processing through individual paths, the outputs are concatenated into a single layer merging the learned features across all inputs. This combined layer is then fed through a sequence of Dense layers before concluding at a single neuron with a `linear' activation function for the output. The final structure of the ML model will be further discussed in the Results chapter.

The compilation and optimization of the model was carried out with an Adam optimizer that was selected for its efficient computation and adaptability in learning rates, helping to manage noisy and complex datasets. The model is trained over 50 epochs as the accuracy gain per epoch plateaus at this point. Epochs of 100 or more did not show significantly more accurate results. With a batch size of 8 we aimed to balance computational efficiency and the required precision for weight adjustments. Any changes will be discussed in the Results chapter. Multiprocessing, including the already mentioned GPU utilization, enhances training efficiency.

\subsection{Parameter Analysis and Model Building}
This chapter analyzes the correnations between various potential input parameters. The Pearson correlation cefficient is the one used in this work.
 
\subsubsection*{Correlation between Parameters}
First, we analyze the correlation between measured parameters to determine the most important input for the ML model. Higher correlation coefficients indicate greater importance in calculating the SKR. Highly correlated values could simplify finding patterns during the training of the Keras model. We examine the correlation coefficients between various key metrics to understand the interdependencies among the parameters and their impact on the QKD system's performance. Table~\ref{tab:correlation_l1} shows an overview of all parameters' correlations of link 1. The correlation between the SKR and the QBER is notably negative at -0.3095 which is indicative of their inverse proportionality. This negative correlation is vital as it suggests that improvements in QBER directly enhance the SKR, providing a clear target for system optimization.\\
Furthermore, a high positive correlation of 0.8861 between SKR and visibility underscores the crucial role of optical visibility in the efficiency of key generation. High visibility correlates with less noise and more distinguishable quantum states, improving the SKR and making it a significant parameter for inclusion in our predictive models.\\
Conversely, correlation between SKR laser power is low, at 0.015220. This low value indicates that laser power has a reduced impact on SKR within our experimental setup, suggesting is is less useful for training our ML model.\\

\begin{table}%[h]
	\centering
	\begin{tabular}{|c|r|r|r|r|r|}
		\hline
 		Parameter 			& SKR 		& QBER 		& Visibility 	& Laserpower 	\\ \hline
		SKR 				& 1.000000 	& -0.309539 & 0.886188 		& 0.015220 		\\ \hline
		QBER 				& -0.309539 & 1.000000 	& -0.235367 	& 0.022756 	\\ \hline
		Visibility 			& 0.886188 	& -0.235367 & 1.000000 		& 0.026892 	\\ \hline
		Laserpower 			& 0.015220 	& 0.022756 	& 0.026892 		& 1.000000 	\\ \hline
	\end{tabular}
	\caption{Correlation coefficients between parameters of link 1}
	\label{tab:correlation_l1}
\end{table}

\begin{table}%[h]
    \centering
    \begin{tabular}{|c|r|r|r|r|r|}
        \hline
        Parameter 			& SKR 		& QBER 		& Visibility 	& Laserpower 	\\ \hline
        SKR          		& 1.000000 	& 0.025285 	& 0.297132 		& -0.004153 	\\ \hline
        QBER             	& 0.025285 	& 1.000000 	& 0.058996 		& 0.002246  \\ \hline
        Visibility       	& 0.297132 	& 0.058996 	& 1.000000 		& 0.010353  \\ \hline
        Laserpower       	& -0.004153 & 0.002246 	& 0.010353 		& 1.000000  \\ \hline
    \end{tabular}
    \caption{Correlation coefficients between parameters of link 2}
    \label{tab:correlation_l2}
\end{table}

The correlation matrices for link 1 (Table~\ref{tab:correlation_l1}) and link 2 (Table~\ref{tab:correlation_l2}) indicate that the relationships among the parameters differ notably between the two links. Link 2 has a much weaker correlation between SKR and QBER at 0.025285, suggesting that QBER has a reduced impact on SKR in this link. Although the relation between QBER and SKR is theoretically inversely proportional, the measured data of Link 2 contradicts this relation.\\
Similarly, the correlation between SKR and visibility in link 1 is at 0.886188, while contrastingly link 2 displays a weaker correlation at 0.297132. This indicates that while visibility still impacts SKR, the effect is reduced compared  to link 1.\\
Furthermore, the correlations between SKR andlaser power isnegligible across both links with coefficients close to zero. Laser power does not significantly influence SKR, affirming its lower relevance in modeling efforts focused on predicting SKR improvements.\\
These disparities in correlation strength between the links emphasize the variability in system conditions or configurations and suggest that models or strategies effective in one link setting are not directly transferrable to another without adjustments. This analysis not only guides the optimal parameter focus for enhancing system performance in each link but also aids in the customization of predictive models for different QKD setups.

We introduce link loss to generalize the ML model and predict different links with one model. The  link loss represents the total losses occurring on the way from Alice to Bob, which is a helpful parameter for the ML prediction if the model is trained by measurement of one link and another link shall be evaluated.

\subsubsection*{Correlations between Current and Past SKRs}

\begin{table}%[h]
	\centering
	\begin{tabular}{|c|c|c|c|c|c|c|}
		\hline
		Parameter   & SKR 	& SKR\_10 & SKR\_20 & SKR\_30 & SKR\_40 & SKR\_50 \\ \hline
		SKR     & 1.000000  & 0.902975    & 0.866732    & 0.845280    & 0.824011    & 0.801321    \\ \hline
		SKR\_10 & 0.902975  & 1.000000    & 0.902146    & 0.866699    & 0.844956    & 0.823900    \\ \hline
		SKR\_20 & 0.866732  & 0.902146    & 1.000000    & 0.901918    & 0.866564    & 0.843683    \\ \hline
		SKR\_30 & 0.845280  & 0.866699    & 0.901918    & 1.000000    & 0.901889    & 0.866255    \\ \hline
		SKR\_40 & 0.824011  & 0.844956    & 0.866564    & 0.901889    & 1.000000    & 0.901287    \\ \hline
		SKR\_50 & 0.801321  & 0.823900    & 0.843683    & 0.866255    & 0.901287    & 1.000000    \\ \hline
	\end{tabular}
	\caption{Correlation coefficients between current and past SKRs of link 1}
	\label{tab:SKR_correlations_l1}
\end{table}

In our QKD system analysis, understanding the temporal dynamics of key generation rates is crucial for optimizing performance and stability. The examination of correlations between the current SKR and its preceding values provides insights into the predictability and consistency of key generation over time. Our analysis, as summarized in the correlation table \ref{tab:SKR_correlations_l1}, highlights that recent past values of SKR have strong correlations with the current SKR, suggesting a pattern or trend that could potentially be leveraged to predict future key rates.\\ 
The correlation coefficients between the current SKR and its values from 10, 20 and 30 minutes ago are 0.902975, 0.866732 and 0.845280 respectively. These high correlation values indicate a strong dependency of the current SKR on its previous values, particularly within a 30-minute window. This temporal dependency is likely due to the inherent stability of the physical and environmental conditions affecting the QKD system within short periods. Factors such as temperature stability, fiber alignment and minimal changes in quantum channel conditions contribute to the consistency of the SKR over such intervals.\\
Given these strong correlations, we analyze the impact of the first three previous values of the SKR (SKR\_10, SKR\_20, and SKR\_30) in our ML models for predicting future SKR values. The inclusion of these parameters can enhance the model's ability to capture the temporal dynamics of the key generation process, providing a more accurate and reliable prediction of key rates.\\

\begin{table}%[h]
	\centering
	\begin{tabular}{|c|r|r|r|r|}
		\hline
		Parameters  & SKR & SKR\_10 & SKR\_20 & SKR\_30 \\ \hline
		SKR     & 1.000000 & 0.251248 & 0.204925 & 0.149124 \\ \hline
		SKR\_10 & 0.251248 & 1.000000 & 0.250553 & 0.206416 \\ \hline
		SKR\_20 & 0.204925 & 0.250553 & 1.000000 & 0.250686 \\ \hline
		SKR\_30 & 0.149124 & 0.206416 & 0.250686 & 1.000000 \\ \hline
	\end{tabular}
	\caption{Correlation coefficients between current and past SKRs of link 2}
	\label{tab:SKR_correlations_l2}
\end{table}

Table~\ref{tab:SKR_correlations_l2} shows the correlations between previous values of the SKR.
Comparing Table~\ref{tab:SKR_correlations_l1} with Table~\ref{tab:SKR_correlations_l2}, we observe a similar phenomenon as that previously described. Link 2, Table~\ref{tab:SKR_correlations_l2}, has qualitatively lower correlations between the SKRs, but the correlation between more recent SKRs is higher than the correlation of older values.

\newpage
\section{Introduction of Measurement Metrics}

To make our models more comparable to each other, we introduce specific errors to establish clear metrics for measuring model differences. This standardization allows us to evaluate and compare the performance of various models on a consistent basis, ensuring that any improvements or regressions in accuracy can be precisely quantified. By defining these errors, we create a robust framework for assessing model effectiveness and identifying areas for further optimization.

The primary evaluation metrics we introduce are the Mean Error (ME), Mean Absolute Error (MAE), Mean Relative Error (MRE) and Mean Squared Error (MSE).

%\subsubsection*{Error (E)}

The Mean Error (ME) represents the mean of the differences between the predicted and actual values. Unlike the AE, it does not take the absolute value of the differences:

\begin{equation}
\text{ME} = \frac{1}{n} \sum_{i=1}^{n} \left( \text{SKR}_{\text{Predicted}}[i] - \text{SKR}_{\text{Measured}}[i] \right)
\end{equation}

This metric is useful for understanding the tendency in the predictions. This error shows if our predictor tends to over- or underpredict for the specific link. 
%\subsubsection*{Absolute Error (AE)}

The Mean Absolute Error (MAE) measures the average magnitude of the errors in a set of predictions without considering their direction. It is the average of the absolute differences between the predicted and actual values:

\begin{equation}
\text{MAE} = \frac{1}{n} \sum_{i=1}^{n} \left| \text{SKR}_{\text{Predicted}}[i] - \text{SKR}_{\text{Measured}}[i] \right|
\end{equation}

%\subsubsection*{Relative Error (RE)}

The Mean Relative Error (MRE) provides a normalized measure of the error, allowing us to understand the error magnitude in relation to the actual values. It is calculated as the average of the absolute differences between predicted and actual values divided by the actual values:

\begin{equation}
\text{MRE} = \frac{1}{n} \sum_{i=1}^{n} \frac{ \text{SKR}_{\text{Predicted}}[i] - \text{SKR}_{\text{Measured}}[i]}{\text{SKR}_{\text{Measured}}[i]}
\end{equation}

The ME, MAE and MRE errors are calculated for the non-scaled data to make the errors more demonstrative for us as they are measured in bits/s.

%\subsubsection*{Mean Squared Error (MSE)}
The Mean Squared Error (MSE) is a commonly used metric for assessing the accuracy of a predictive model. It measures the average of the squares of the errors—that is, the average squared difference between the predicted and actual values. The MSE is calculated as follows:

\begin{equation}
\text{MSE} = \frac{1}{n} \sum_{i=1}^{n} (\text{SKR}_{\text{Measured}}[i] - \text{SKR}_{\text{Predicted}}[i])^2
\end{equation}

Where $\text{SKR}_{\text{Measured}}$ represents the actual values stemming from the CSV files and the $\text{SKR}_{\text{Predicted}}$ represents the values predicted by the ML model. The MSE is calculated for the scaled data which is particularly useful for detecting outliers and ensuring the model works well on the whole dataset.
Besides the rest of the errors (ME, MAE, MRE), the MSE error measurement is calculated with the scaled data, which is useful for comparing errors across different scales and links. This makes it easier to compare links with higher SKRs to links with lower SKRs.

These metrics collectively provide a comprehensive assessment of the prediction accuracy of the model and are crucial for evaluating the performance of our predictive models in the context of link quality and reliability.

\chapter{Results}

In this chapter, we present and analyze the empirical data collected from our QKD data analysis. The results are systematically organized to address the primary research objectives, focusing on how accurate theoretical QKD models can model a real-world QKD network.

We use the following convention to differentiate between the results of theoretical COW and ML models. If results from the theoretical COW model class are presented, the corresponding data is ``calculate", ``estimated" and ``fitted". ``Predicted" and ``trained" are used in reference to data from the ML model.

\section{Theoretical COW Model Calculations and Fitting}

This section discusses the implemented COW model based on theoretical knowledge about COW QKD networks. How well it performs in estimating the SKR of the QKD system and if this model can be improved further.

\begin{figure}[h]%[h!]
  \begin{center}
    \includegraphics[width=0.49\textwidth]{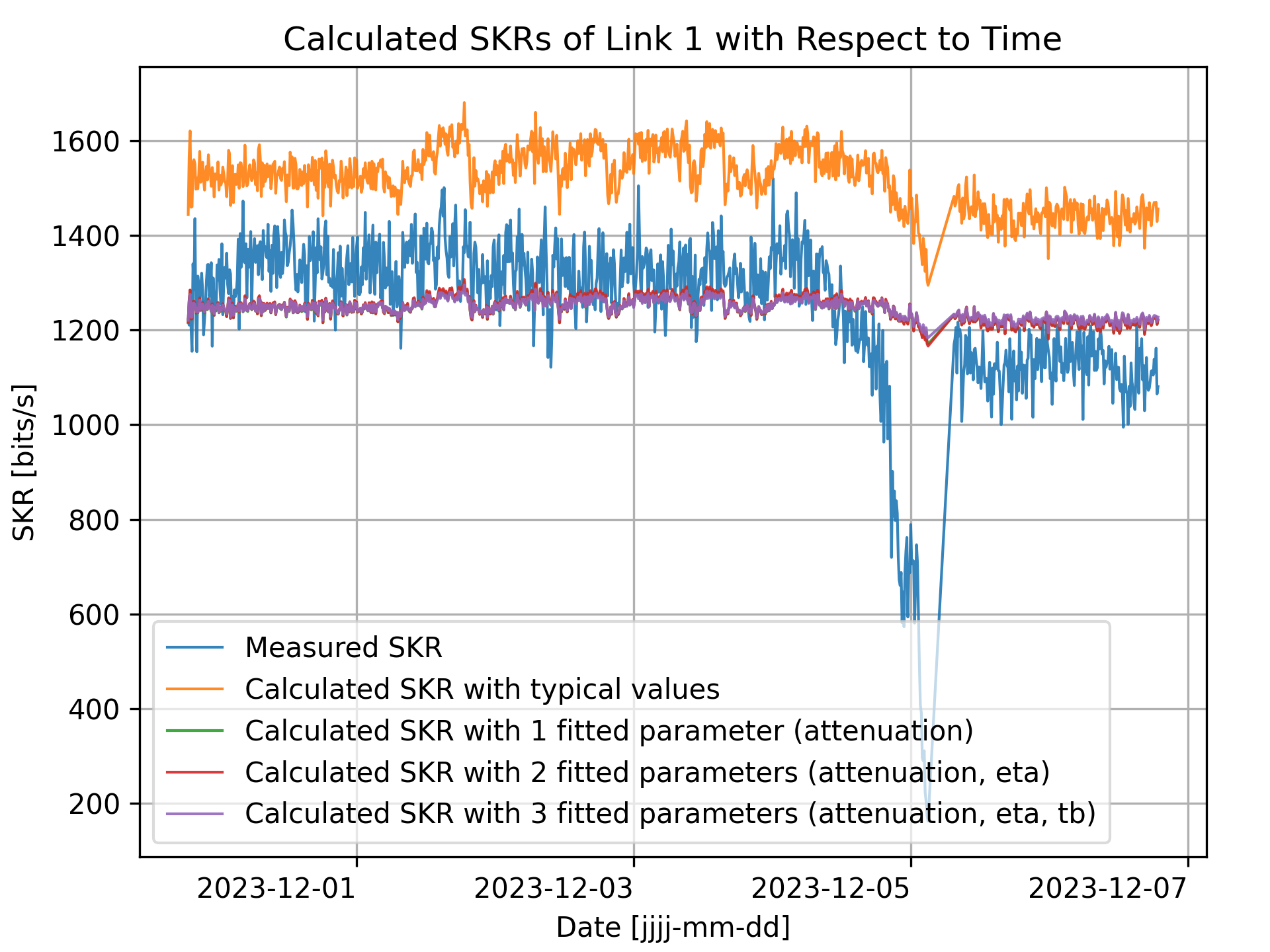}
    \includegraphics[width=0.49\textwidth]{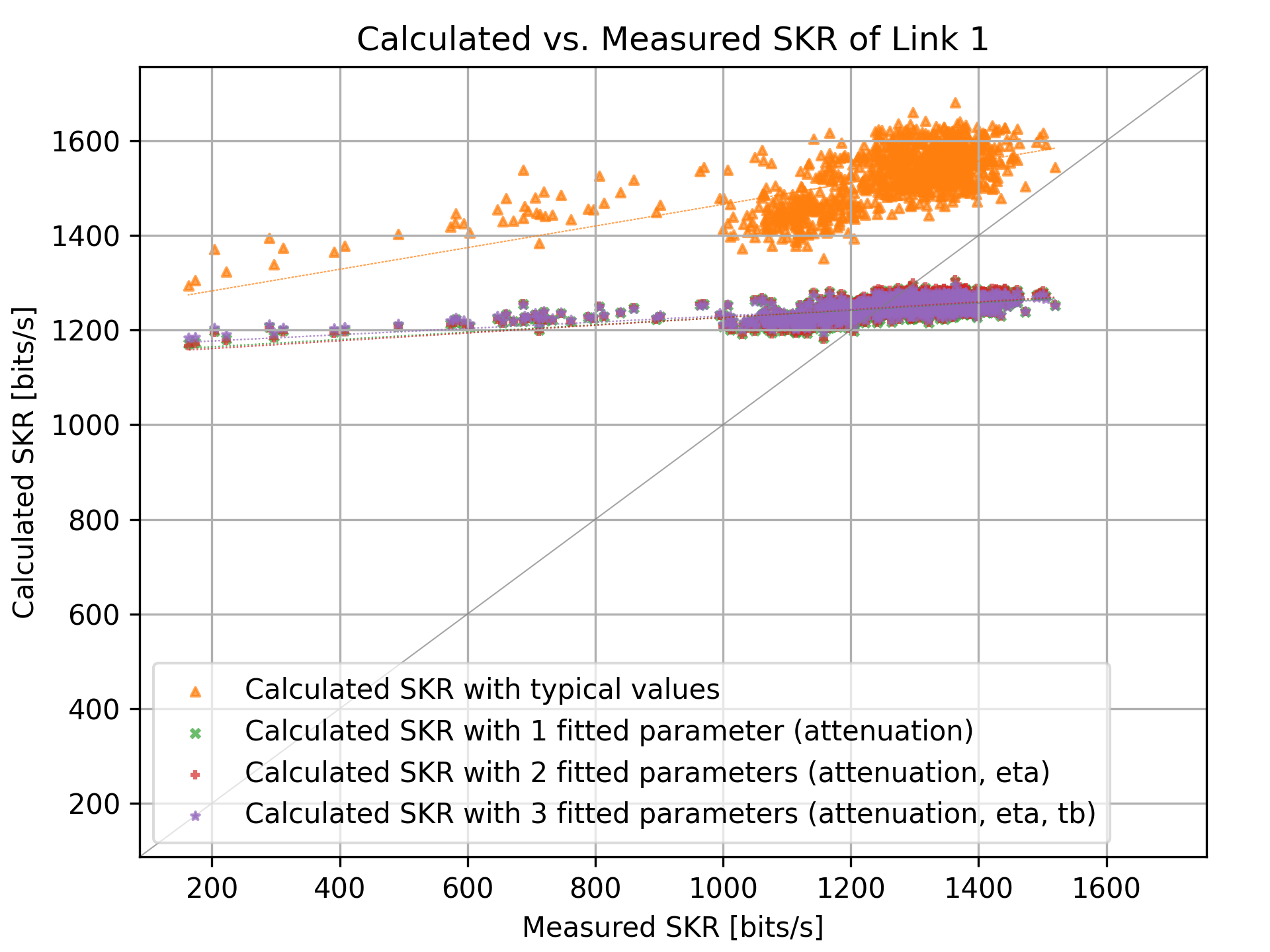}
    \caption{Calculated SKR of link 1 before and after curve fitting with respect to time and to the measured values}
    \label{fig:curvefit_l1}
  \end{center}
\end{figure}

\begin{figure}[h!]%[h!]
  \begin{center}
    \includegraphics[width=0.49\textwidth]{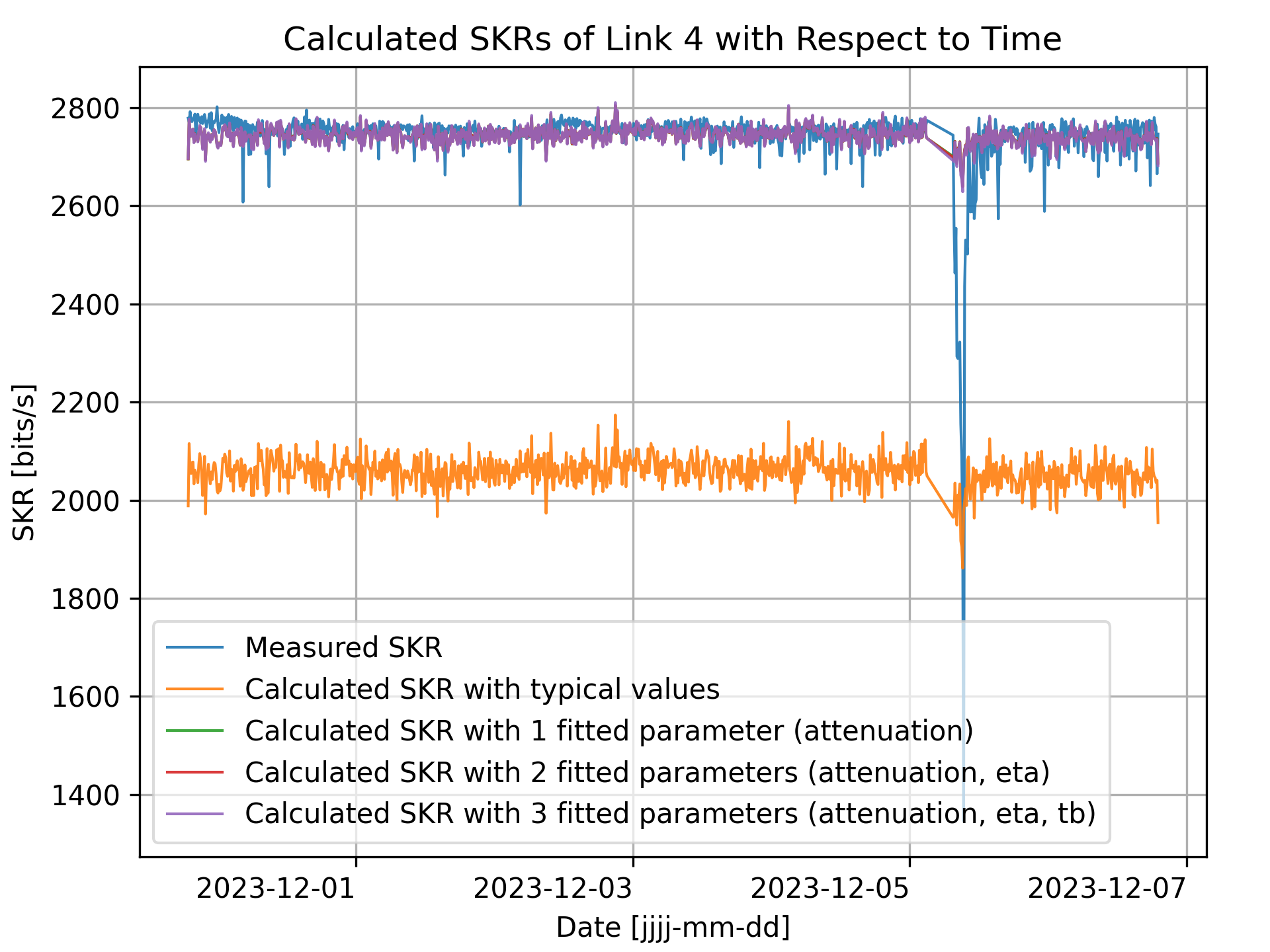}
    \includegraphics[width=0.49\textwidth]{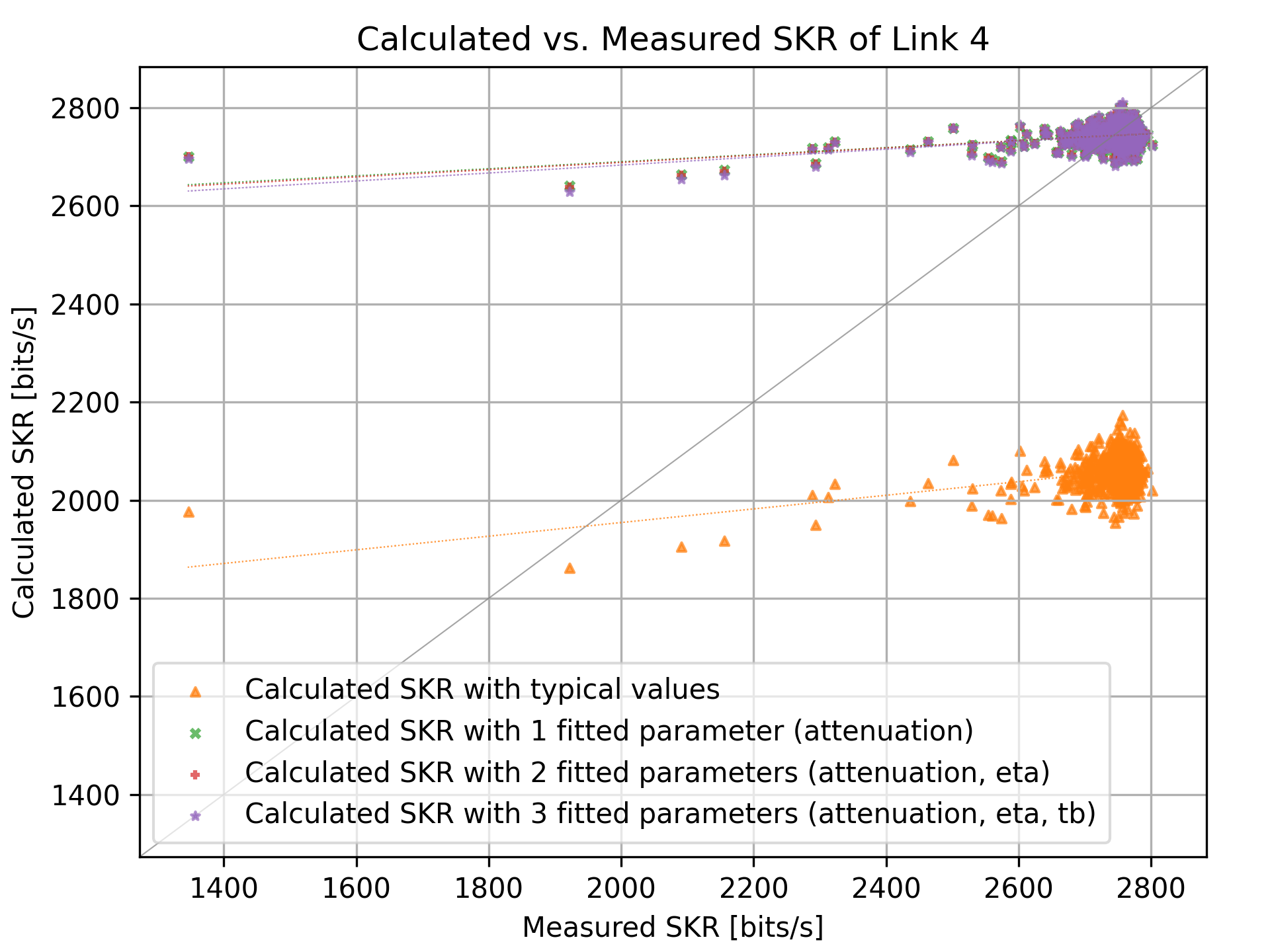}
    \caption{Calculated SKR of link 4 before and after curve fitting with respect to time and to the measured values}
    \label{fig:curvefit_l4}
  \end{center}
\end{figure}

\begin{figure}[h!]%[h!]
  \begin{center}
    \includegraphics[width=0.49\textwidth]{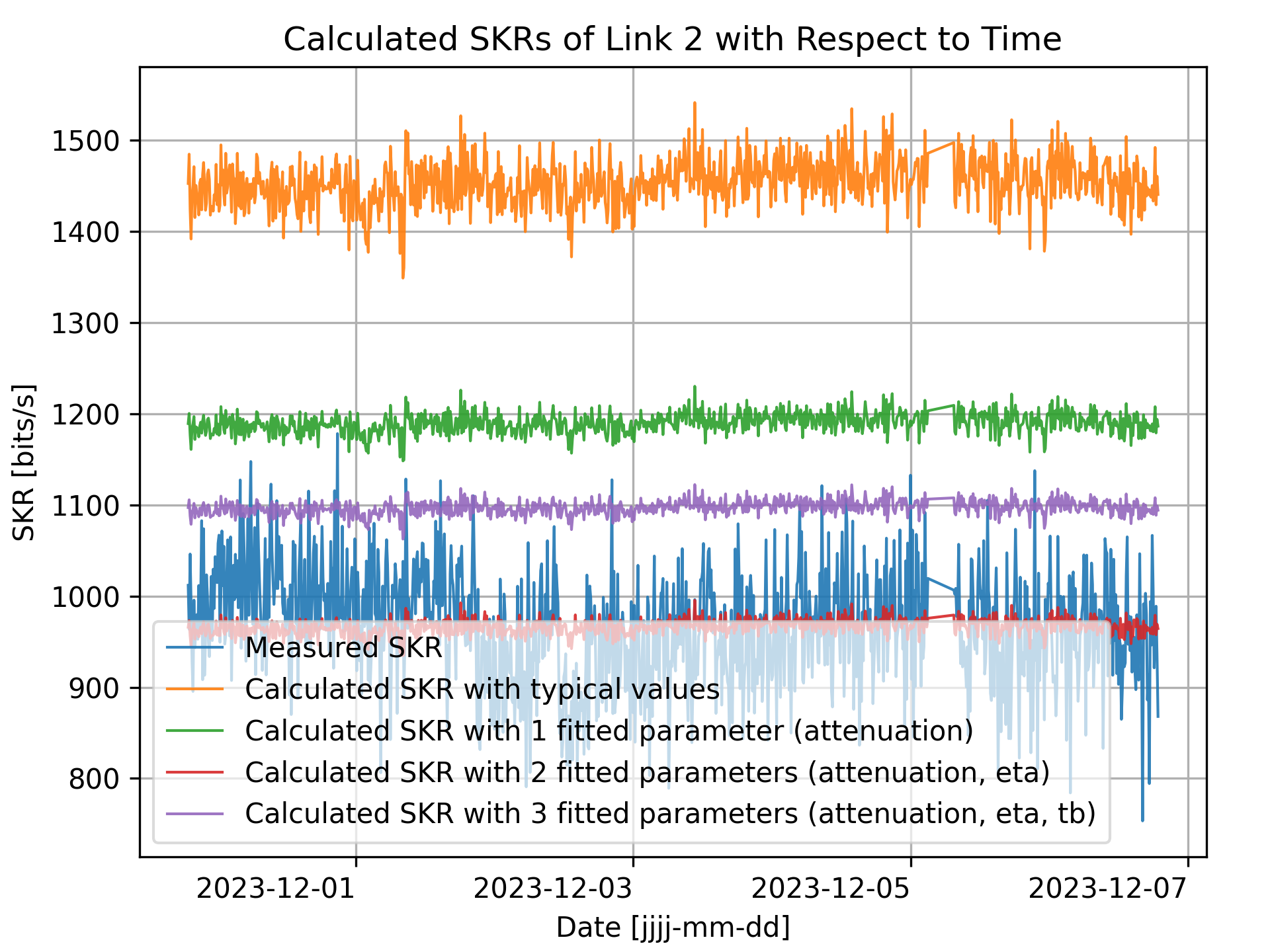}
    \includegraphics[width=0.49\textwidth]{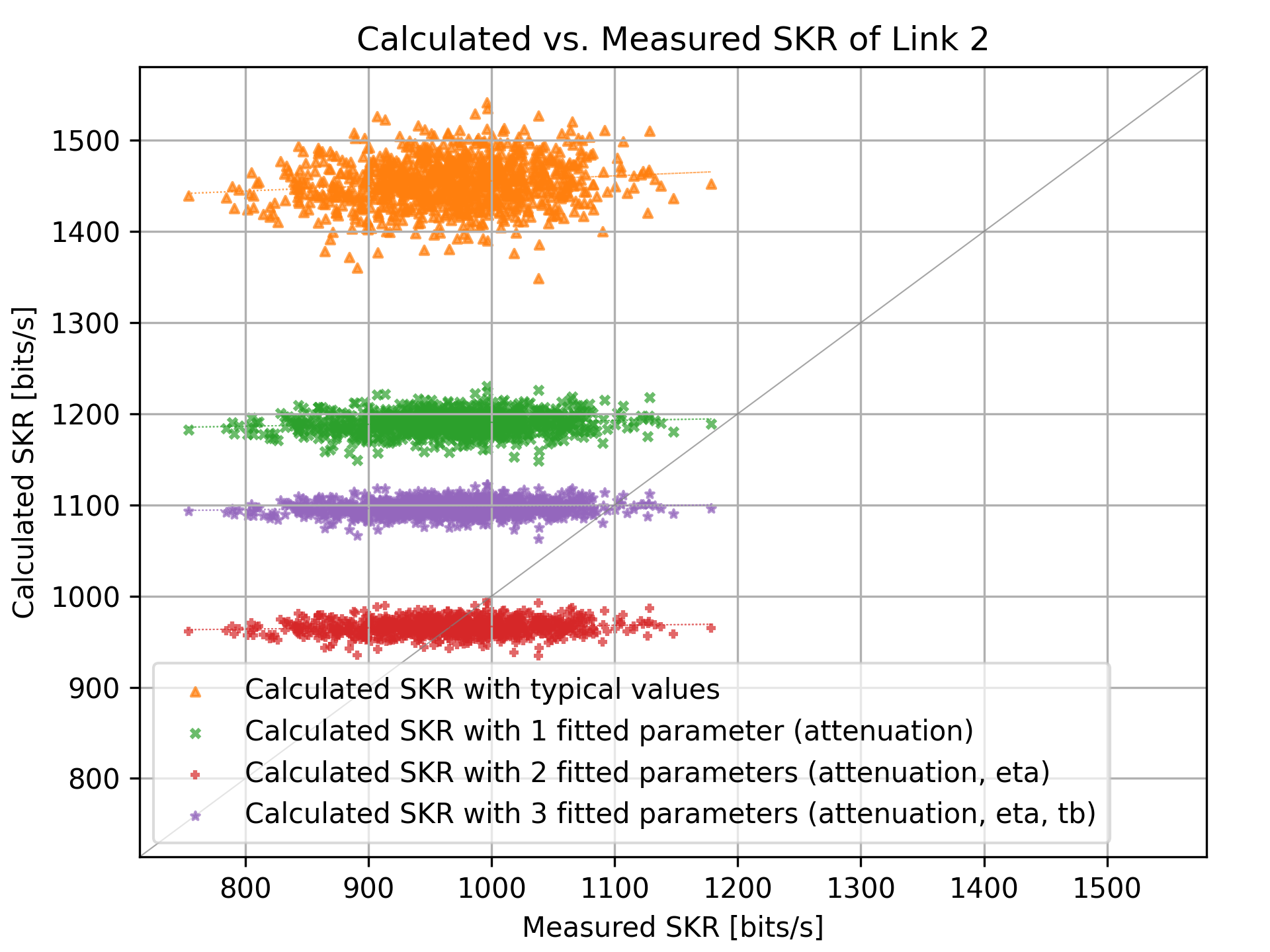}
    \caption{Comparison of measured SKR with the calculated SKR of link 2 before and after curve fitting in two different plot types}
    \label{fig:curvefit_l2}
  \end{center}
\end{figure}

Figure~\ref{fig:curvefit_l1} consists of two subplots: the left subplot illustrates the measured and calculated SKR over time, while the right subplot is a scatterplot comparing the calculated SKR (y-axis) to the measured SKR (x-axis).\\
In the left subplot, five curves are presented: the first represents the measured data of our COW network link 1; the second represents the initial COW model with predefined parameter values; and the last three represent the COW model after fitting the parameters to the measured data. The legend differentiates these curves and the states of the COW model. Each COW model fitted a different number of parameters to show its influence on the SKR calculation: The first model only fitted the attenuation; the second also fitted the eta; and the third added the loss of Bob's components to the fitting process. Howerver Figure~\ref{fig:curvefit_l1} and Figure~\ref{fig:curvefit_l4} do not show all three COW model fits, as all three fits are nearly identical. Only Figure~\ref{fig:curvefit_l2} shows these differences which is discussed further down in this section.\\
The right subplot further analyzes the accuracy of the COW model by plotting the calculated SKR against the measured SKR. Ideally, points should lie along the diagonal line (y=x) if the model perfectly matched the measurements.
Figure~\ref{fig:curvefit_l1} demonstrates that fitting the COW parameters significantly improves the model's representation of the network link performance. The model with predefined values mostly overestimates the SKR since the channel and real-world conditions reduce the SKR.\\
Figure~\ref{fig:curvefit_l1} also shows a heavy drop in the SKR on day 5, possibly due to a system malfunction. Since the COW model and its equations are meant to represent a QKD system during regular operation, it is acceptable that the model does not provide accurate SKR calculations during this non-save state.\\
Figure~\ref{fig:curvefit_l4} shows the same behavior as in Figure~\ref{fig:curvefit_l1} except that during this comparison the model deliberatly underestimates the SKR. This is probably because link 4 performs better than we estimated with the predefined parameter values.

% Todo: table with errors?

Figure~\ref{fig:curvefit_l2} shows the same graphical comparison of the SKR as Figures~\ref{fig:curvefit_l1} and \ref{fig:curvefit_l4}, but in this experiment different behaviour is revealed. Unlike link 1 and link 4 (where the three fitting curves where almost identical), link 2 improves the calculated SKR when the fitted parameters are increased from 1 to 2. These differences in parameter fitting behaviors across the links as shown in figures~\ref{fig:curvefit_l1},~\ref{fig:curvefit_l4} and ~\ref{fig:curvefit_l2} highlight the variability in system conditions or configurations. This indicates that models or strategies successful in one link setting may not be directly applicable to another without modifications.

Despite the accuracy improvements we gain from the COW model fitting, the model's accuracy remains insufficient. This prompts exploring a more complex solution involving an ML model.

%\subsubsection*{Plots with theoretical maximal SKR}

%Todo: show and describe the Plots with the maximal possible SKR

%\subsubsection*{Possion Distribution of the Average Photons per Pulse}

%Todo: show and describe the Poisson distribution of $\mu$ and/or compare laserpower and $\mu$ 

\newpage
\section{Minimizing and Optimizing the ML Model} 
This section details the optimization of the ML model to enhance its performance in SKR prediction.

\subsection{Simplified ML Model}

One approach we take to optimizing the ML model is to significantly reduce the complexity of the ML model and match its input layer sizes to the size of our data (1, 1, 1, 3). While this simplification aimes to streamline the model, it results in problematic errors in data measurements because the model could not adequately evaluate all properties' influences on the SKR. Consequently, this simplified model proves suboptimal, highlighting the need for a more sophisticated approach.

\subsection{Optimized ML Model}

To overcome the limitations of the simplified model, we implement several optimization strategies focusing on improving accuracy and generalization. First, we adjust the input layer sizes to better capture the complexity of the data, which results in a more accurate representation of the input parameters.
We then optimize the number of dense (hidden) layers to balance model complexity and performance. This enhancement allows the model to learn more effectively from the data. Additionally, we introduce a learning rate scheduler that scales down the learning rate with increasing epochs, ensuring the model converges more efficiently.
To further enhance the model's robustness, we implement early stopping. This technique halts the training process when the test error ceases to improve after a certain number of epochs, thus preventing overfitting and improving generalization.
The optimized model features multiple branches, each handling different input parameters. For example, each branch starts with a dense layer of 64 units with `ReLU' activation, followed by another dense layer of size 16 perceptron units. The outputs from these branches are then concatenated and passed through a series of dense layers, refining the data through layers with 64, 128, 32 and 8 units respectively, all using `ReLU' activation. The final output layer employs a `linear' activation function to produce the prediction.
The model training utilizes the Adam optimizer with an initial learning rate 0.001. The learning rate scheduler reduces the learning rate by 1\% for each epoch starting after the 15th epoch, while early stopping monitors the validation loss and ceases training if no improvement is seen after 15 epochs, restoring the best weights.
This optimized configuration significantly reduces training and testing errors, ensuring accurate predictions of QKD system parameters while maintaining generalization across different links. Detailed results and performance plots of the model are presented in this chapter.
It is also worth mentioning that we closely monitor the training and testing errors to ensure the ML model does not overfit. As these errors are not critical to our main findings, we have chosen not to include them here.

\newpage
\section{ML Model Predictions and Performance}
%Presentation of the ML Model's Predictions

This section discusses the results of the ML model, our next-level approach for data prediction.
In this section various approaches to find the best predicting model are discussed. At first there is a discussion about the usefulness of certain model input parameters. Then there is a comparison of different approches for predicting unknown links, e.g. predicting link that are not in the training set. At last the final optimized ML model is presented.

All these models, whether it be a single link prediction or a multi link prediction, use a 80/20 train/test split for their training links. This means that for the models where 2 or more links were used for training and other links were used for evaluation, only 80\% of the training links' data is used for the actual ML model training.  

%\subsection{Presentation of Data Analysis and Model Predictions}

\subsection{Parameter Analysis and Model Building Approaches}

The methodology stated that parameters with a stronger correlation to the SKR would more likely influence the data prediction if used for the ML model training. This section verifies the stated assumptions with the according results.

\begin{figure}[h]%[h!]
  \begin{center}
    \includegraphics[width=0.49\textwidth]{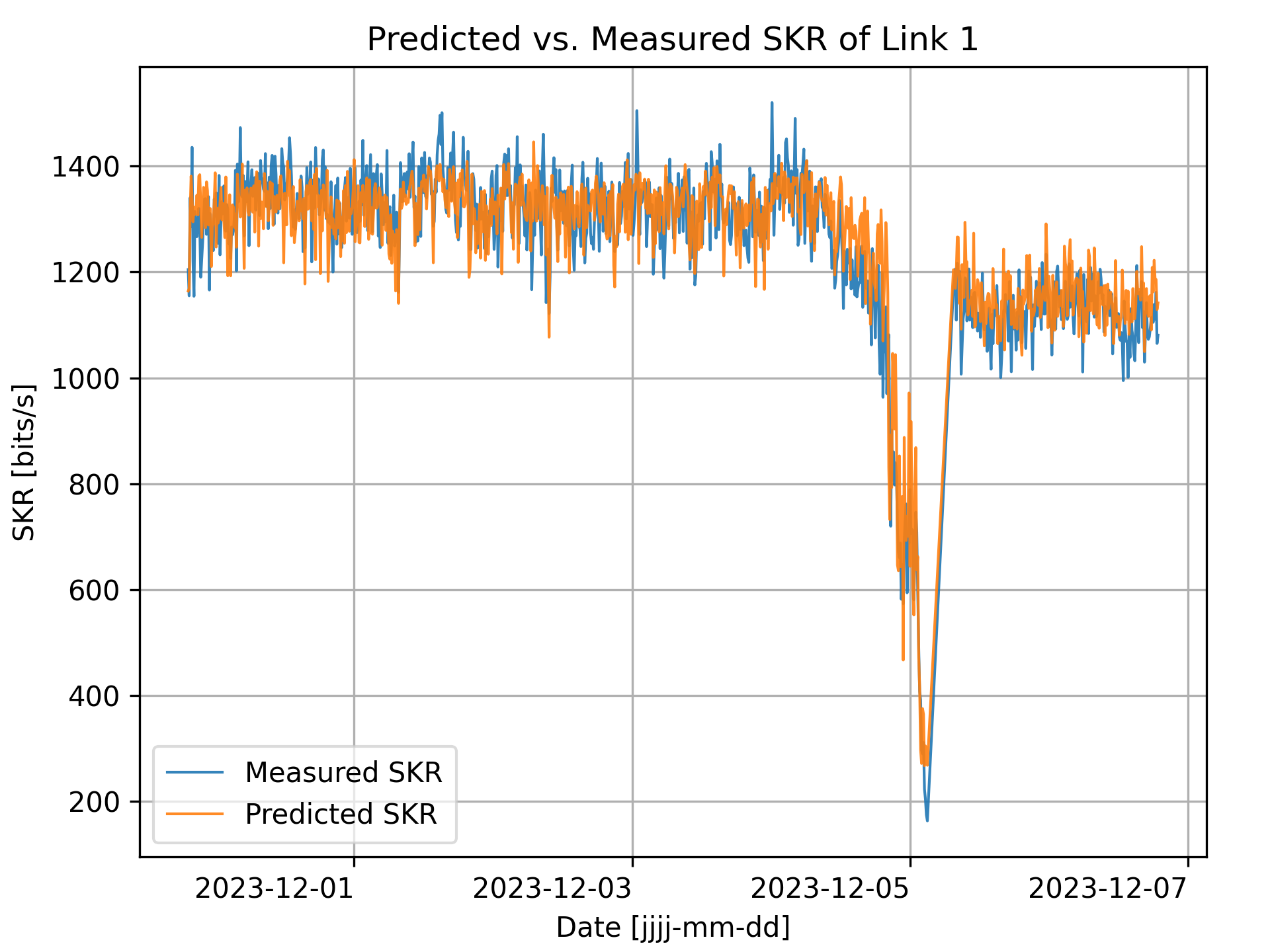}
    \includegraphics[width=0.49\textwidth]{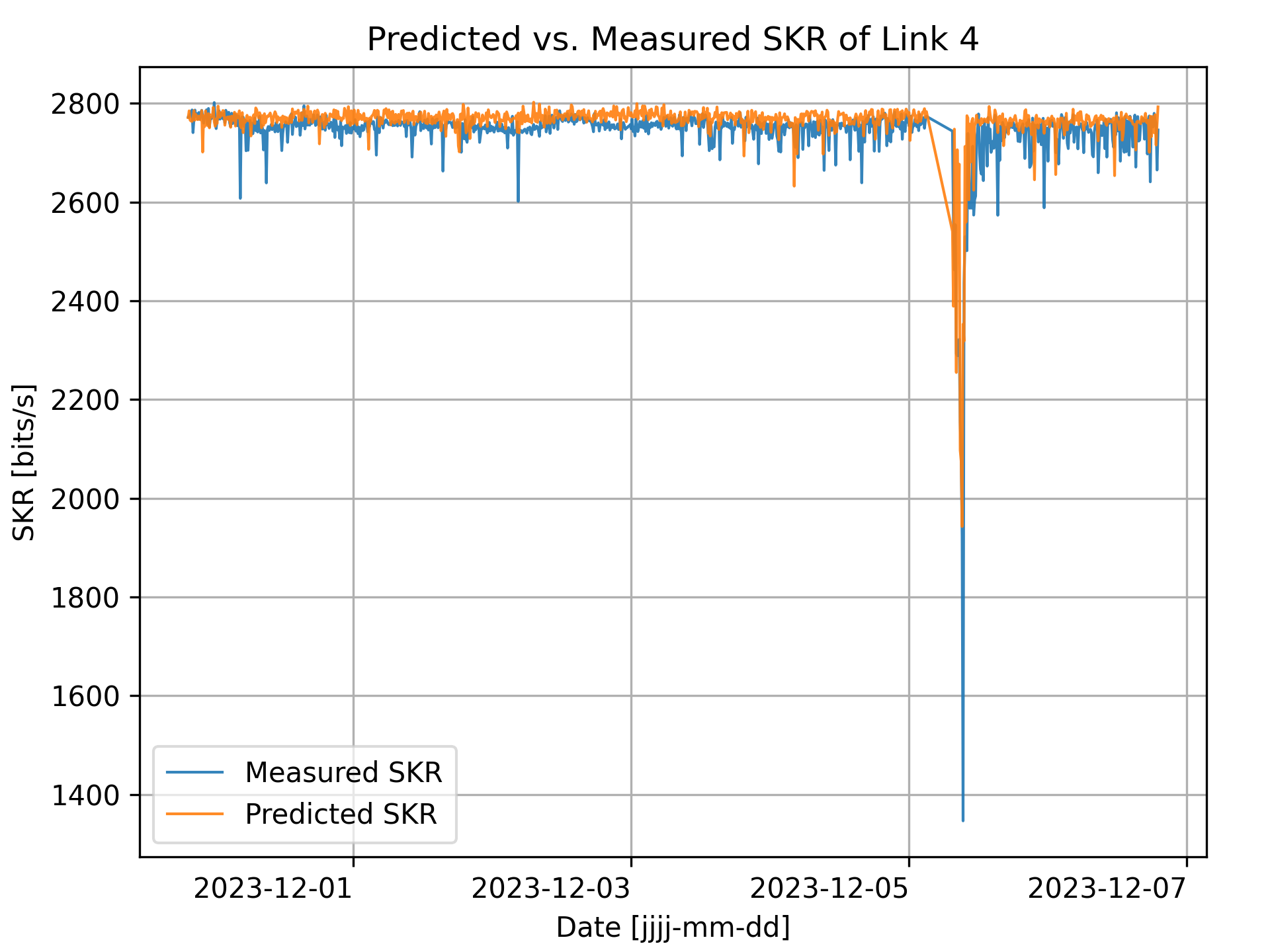}
    \caption{Comparison of measured SKR with the predicted SKR of link 1 and link 4}
    \label{fig:testpredlink14}
  \end{center}
\end{figure}

Figure~\ref{fig:testpredlink14} consists of two subplots that both illustrate the measured and calculated SKR over time. Each subplot shows a separate QKD link. For each link, an individual ML model is trained and evaluated. The predictions made by the trained ML model are shown in this figure. To demonstrate that the model can accurately predict the SKR over the appropriate time frame, we have provided a plot of SKR over time. For future visualizations, we will use scatter plots more frequently, as this plot type conveys the information more clearly.

\begin{figure}[h]%[h!]
  \begin{center}
    \includegraphics[width=0.49\textwidth]{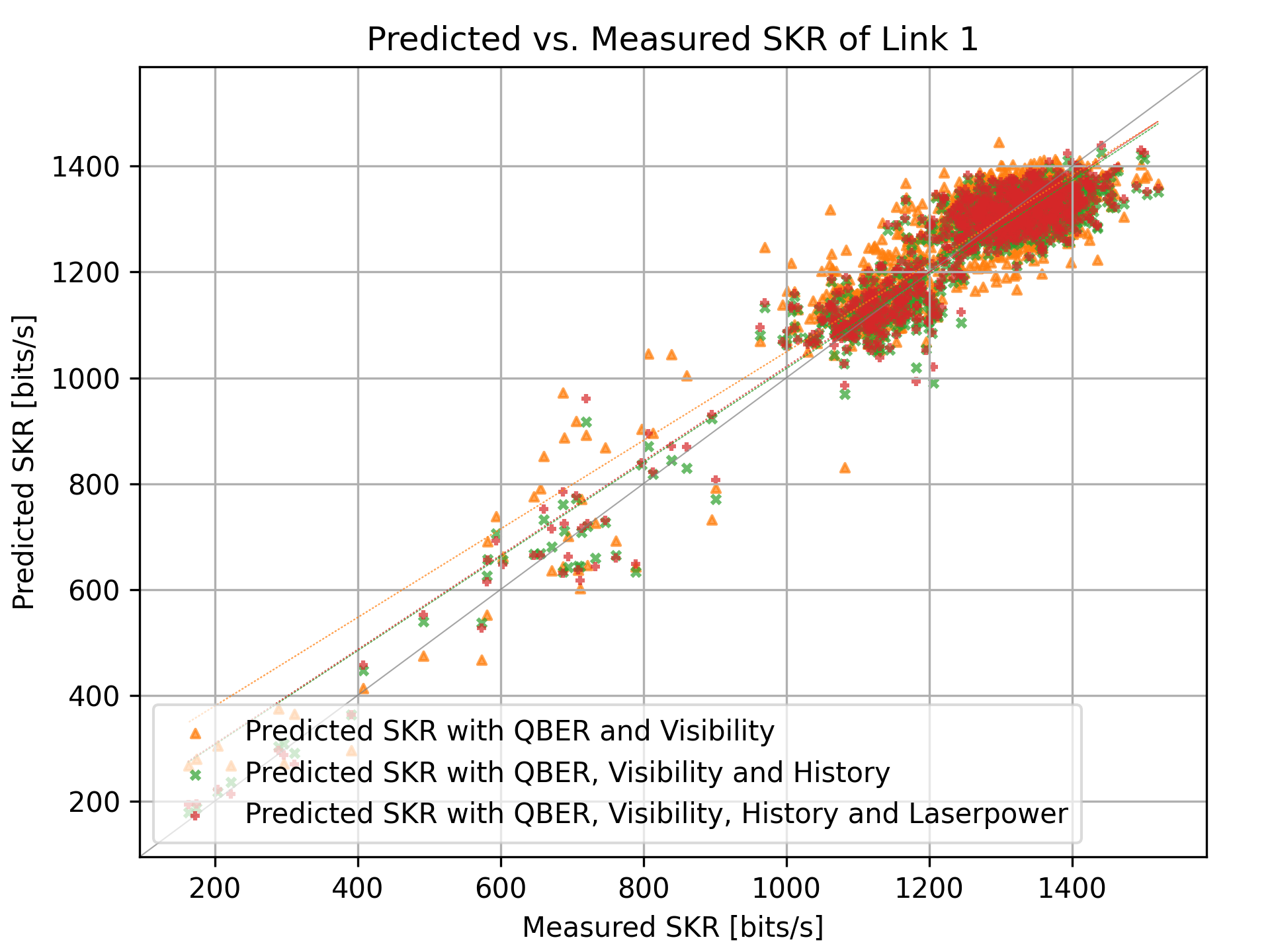}
    \includegraphics[width=0.49\textwidth]{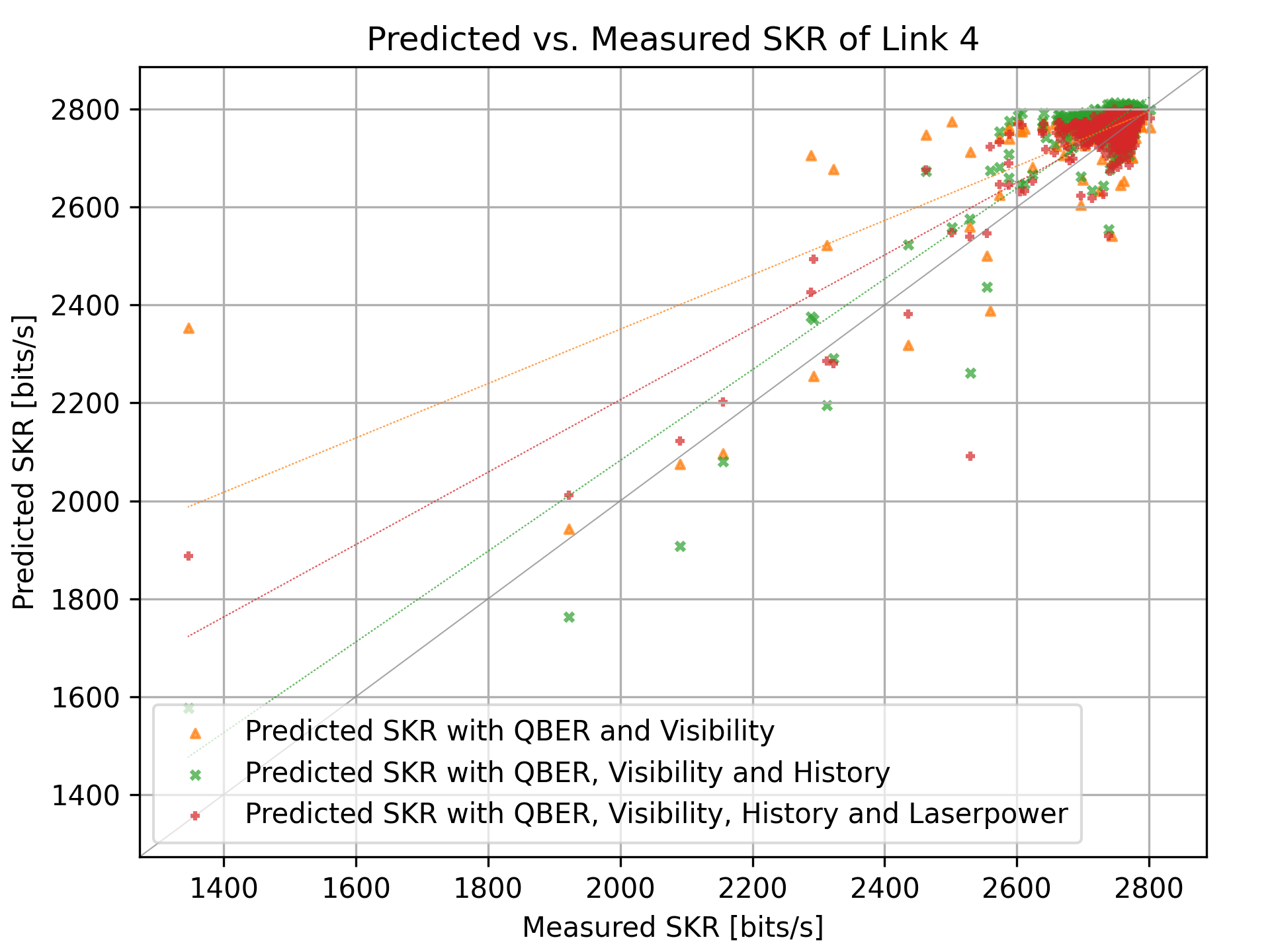}
    \caption{Comparison of measured SKR with the predicted SKR of link 1 and 4 with different training parameters as input data}
    \label{fig:testpredlink14inputparams}
  \end{center}
\end{figure}

Figure~\ref{fig:testpredlink14inputparams} shows the measured and predicted key rates on the x- and y-axis.
%The models used for predictions are the same for both links. 
The legend indicates the input parameters used to train and evaluate the ML models, highlighting that the different colors within one plot (for each link) indicate slightly different models. While one model only utilizes QBER and the visibility as input parameters for the SKR prediction, the second model also includes historical values in its training. The third model additionally uses the laserpower as an input parameter.

\begin{table}%[h]
	\centering
	\begin{tabular}{|l|r|r|r|r|}
		\hline
% 		Link 1					& MSE 		& RE		& AE 		& E  	\\
%        \diagbox[height=1cm]{Input Parameters}{Errors}  &
		&
        \multicolumn{1}{c|}{ME} &
        \multicolumn{1}{c|}{MAE} &
        \multicolumn{1}{c|}{MRE} &
        \multicolumn{1}{c|}{MSE} \\
 		\hline
		QBER, visibility 				& 9.104899	& 54.889449	& 0.012483	& 0.002613 \\
		\hline
		QBER, visibility, history 		& -9.603267	& 43.733839	& -0.005283	& 0.001658		\\
		\hline
		QBER, visibility, history, laserpower & -5.450856 & 43.713529	& -0.001959	& 0.001660	\\
		\hline
	\end{tabular}
	\caption{Comparison of errors of link 1 for different training parameters}
	\label{tab:errors_l1}
\end{table}

Table~\ref{tab:errors_l1} shows the different measurement errors in the horizontal direction and the input parameters for the ML models in the vertical dimension. This table details the influence of various input parameters on the ML models' accuracy.

Figure~\ref{fig:testpredlink14inputparams} and Table~\ref{tab:errors_l1} show an error improvement when we add the historical SKR values are included in the prediction model. Additionally, they demonstrate that adding the laserpower to the input parameters does not much impact on forecasting or even negatively incluencing the estimation negatively, as shown in Figure~\ref{fig:testpredlink14inputparams}, where the line of best fit is most optimal with input values QBER, visibility and history.

\newpage
\subsection{Generalized ML Model Approaches}
This section explores various strategies for developing generalized ML models to ensure robust and accurate predictions across diverse QKD system configurations.

\subsubsection*{Interpolation vs. Extrapolation Approach}

The next step after single link predictions is to predict the performance of all links in our network using an ML model trained on data from two links. From this point forward, we include the link loss as a parameter in the ML model so that the model can differentiate more easily between links and make better predictions for links with lower or higher link losses. To simplify the analysis, we initially use the link distances as a rough estimation for the link losses since the relative link loss values between the links are more important than their absolute values. This means the ML model is trained with data from two links, including their respective link losses. 

To determine the training method, we evaluate two approaches: Interpolation and extrapolation.
Interpolation involves ensuring all evaluation links are withing a comprehensive link loss range for the training of our models. This approach assumes that the training data adequately represents the entire range of possible link conditions, leading to more accurate predictions within this range. By evaluating our model through the interpolation approach, we ensure that our model can reliably predict the behavior of QKD systems for any link condition encountered within the predefined range.

Extrapolation involves training our model with two links on a certain link loss range to ensure that some links from the test set fall outside the link loss range covered by the training data. The case of links being outside the link loss range might occur in field tests when our model is tested on links that have not been covered by our training data yet due to the absence of such good or bad links in the training data.
Comparing the results from both interpolation and extrapolation allows us to identify the most robust training strategy for diverse real-world scenarios and potential limitations.

\begin{figure}[h]%[h!]
  \begin{center}
    \includegraphics[width=0.49\textwidth]{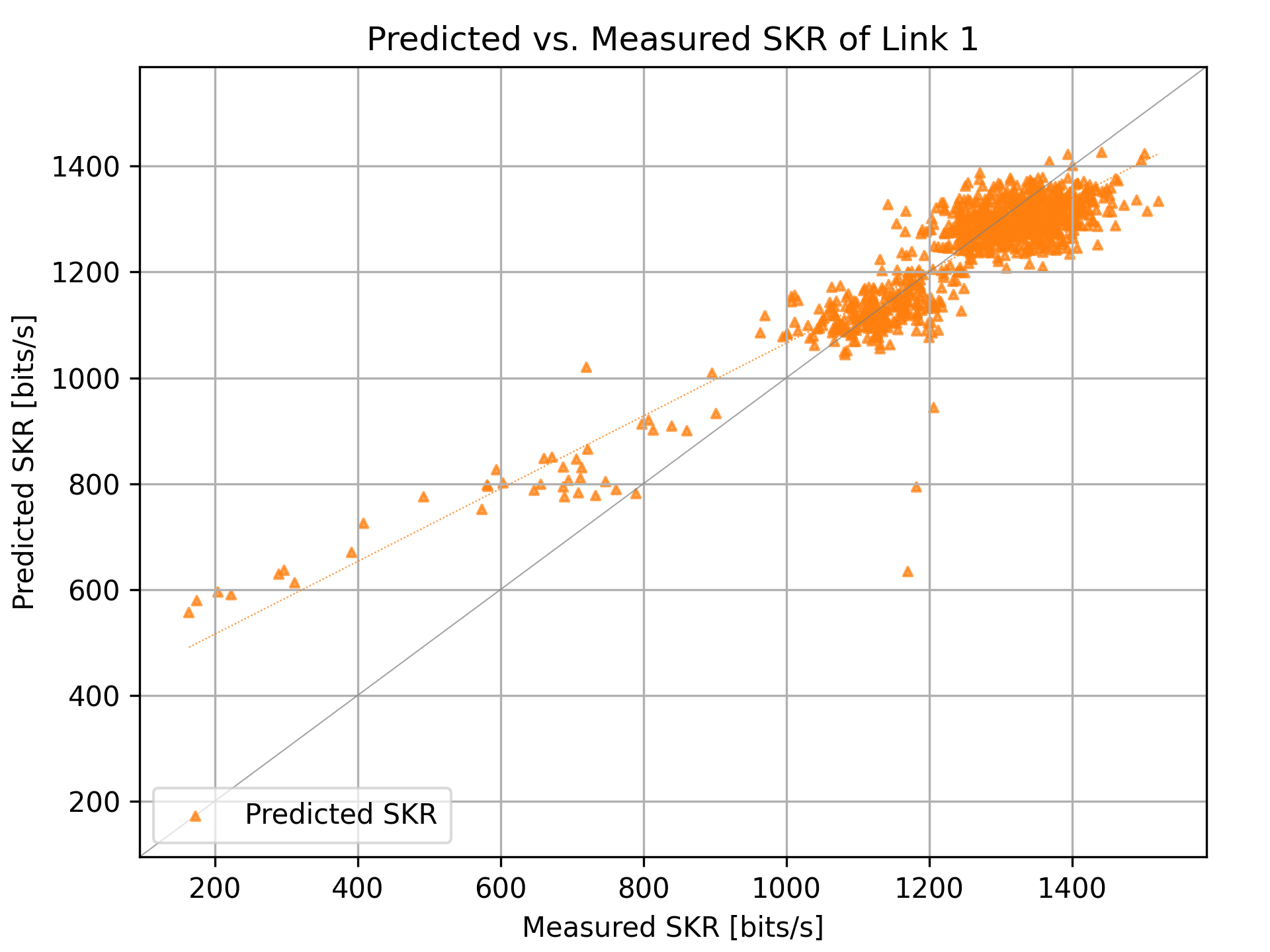}
    \includegraphics[width=0.49\textwidth]{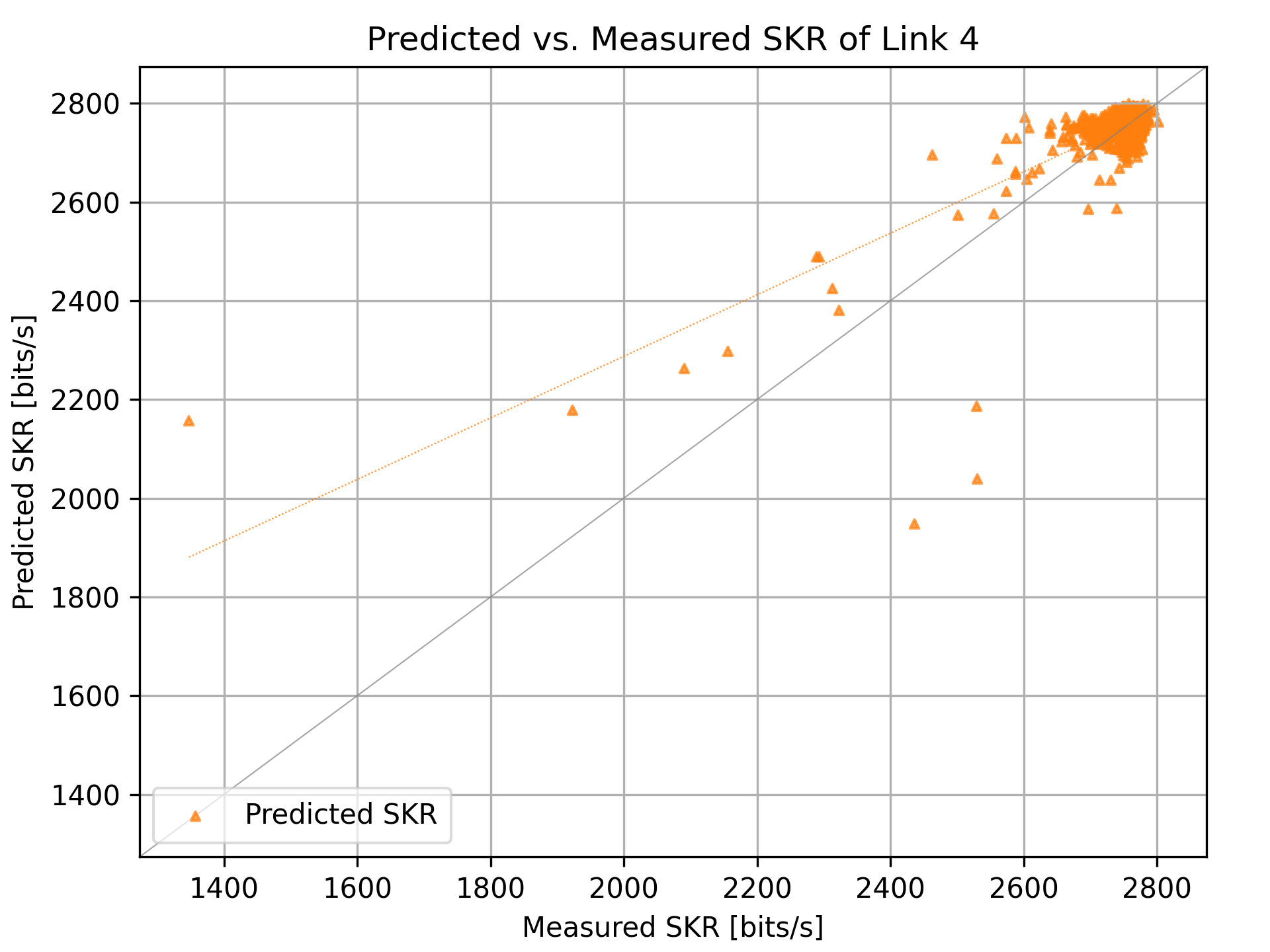}
    \caption{Comparing measured and predicted SKR for link 1 and 4. Interpolating the results for link 1 when training model with link 4}
    \label{fig:interpolation42}
  \end{center}
\end{figure}

Figure~\ref{fig:interpolation42} consists of two subplots comparing the calculated SKR (y-axis) to the measured SKR (x-axis). Both subplots are scatterplots. The left subplot displays the results for link 1 which has been included in the test set, while the right subplot shows the results for link 4, which is part of the training set. In this interpolation example, the ML model is trained with link 4 having a link loss of 20 and link 2 having a link loss of 57. The test link 1 has a link loss of 46. In the left subplot, the points represent the calculated and measured SKR for link 1. Despite being in the test set, the predictions remain relatively accurate because link 1's link loss falls within the spectrum of the training links' link losses. The right subplot shows the comparison for link 4, and since link 4 was part of the training set, the model predicts the SKR quite accurately, with points closely aligning with the diagonal line (y=x).

\begin{figure}%[h]%[h!]
  \begin{center}
    \includegraphics[width=0.49\textwidth]{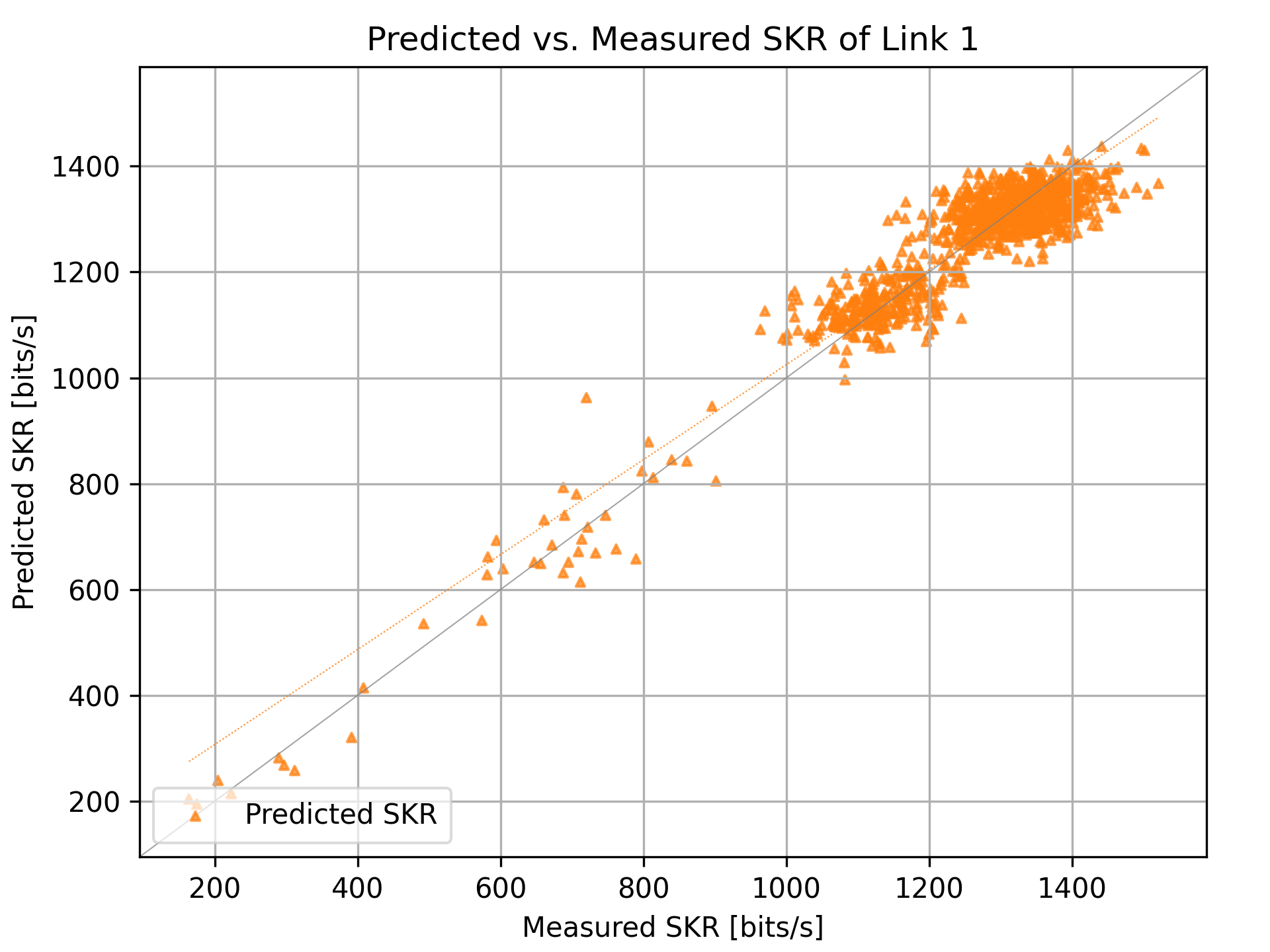}
    \includegraphics[width=0.49\textwidth]{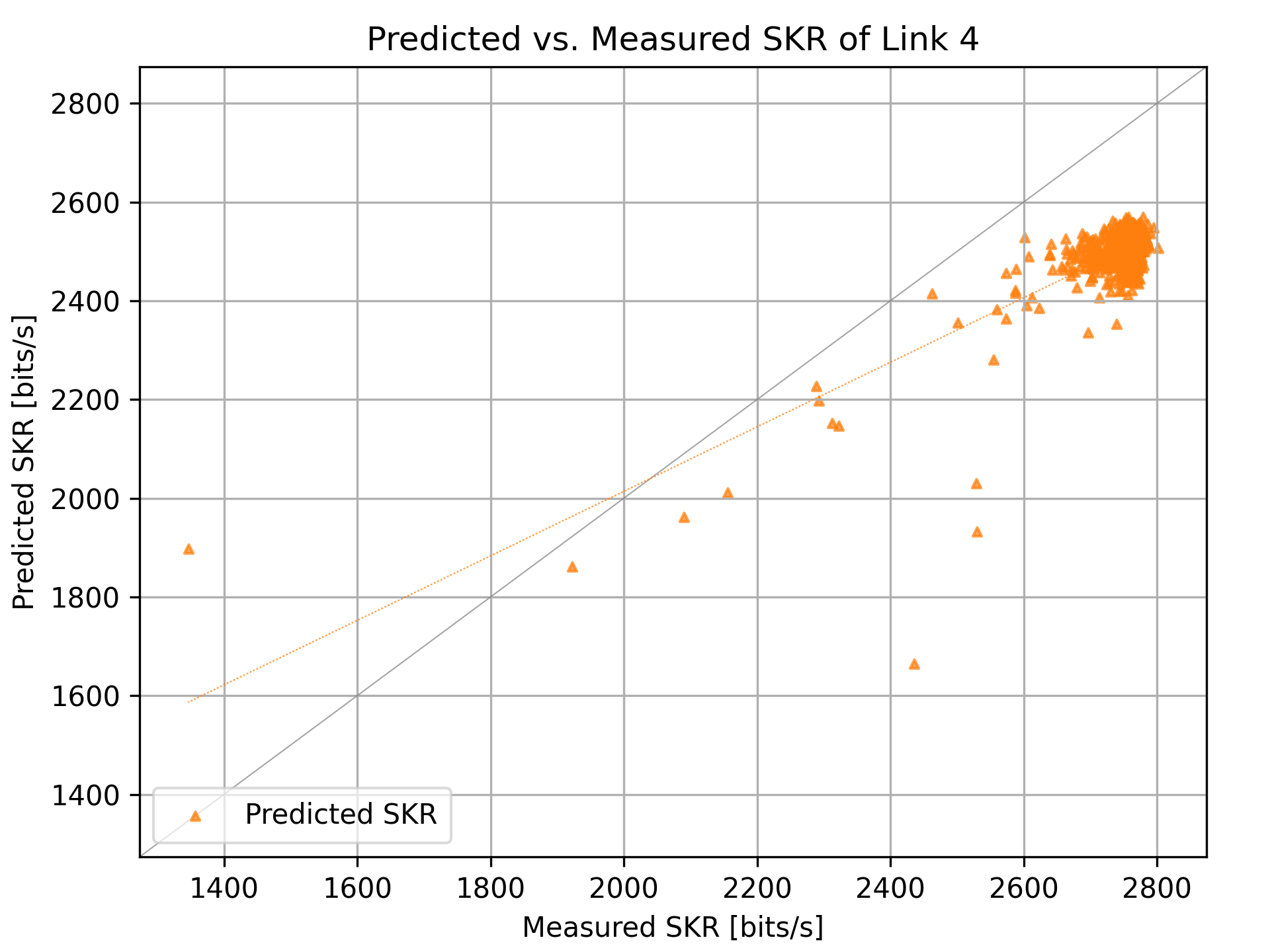}
    \caption{Comparing measured and predicted SKR for link 1 and 4. Extrapolating the results for link 4 when training model with link 1}
    \label{fig:extrapolation12}
  \end{center}
\end{figure}

Figure~\ref{fig:extrapolation12} shows the same plots, but link 1 is included in the training set this time, while link 4 is part of the test/evaluation set. In this extrapolation example, the model is trained with link 1 having a link loss of 46 and link 2 having a link loss of 57. The test link 1 has a link loss of 20. In the left subplot we see the prediction of link 1, which is slightly better since the ML model is trained with link 1 this time. In the plot presenting link 4, we observe that the predictions are worse this time as link 4 is neither included in the training set nor is the link loss inside the training link losses. The right subplot shows the same comparison for link 4. Here, the points deviate more from the diagonal line (y=x), indicating that the model's predictions are less accurate. This discrepancy arises because link 4's link loss lies outside the range of the training links' link losses, highlighting the limitations of the extrapolation approach.
\begin{table}%[h]
	\centering
	\begin{tabular}{|l|r|r|r|r|}
		\hline
		&
        \multicolumn{1}{c|}{ME} &
        \multicolumn{1}{c|}{MAE} &
        \multicolumn{1}{c|}{MRE} &
        \multicolumn{1}{c|}{MSE} \\
        \hline
		Interpolation train error, link 4	& 7.858533	& 23.630816	& 0.003321	& 0.000341 \\
 		\hline
		Extrapolation train error, link 1 	& -5.356389	& 44.679121	& -0.000257	& 0.000466 \\
		\hline
		Interpolation test error, link 1	& -10.9515	& 53.369622	& 0.007223	& 0.000813 \\
		\hline
		Extrapolation test error, link 4 	& -330.587135 & 331.537441 & -0.120235 & 0.016024 \\
		\hline
	\end{tabular}
	\caption{Comparing training and test errors of the interpolated and the extrapolated ML model}
	\label{tab:errors_extra_intra}
\end{table}
Table~\ref{tab:errors_extra_intra} shows the different measurement errors in the horizontal direction with the links and models in the vertical dimension. This table details the accuracy differences between interpolation and extrapolation links. For both links, the training errors are lower than the test errors. However, the model that extrapolates the test link has the least accurate predictions.

Figure~\ref{fig:extrapolation12} and Table~\ref{tab:errors_extra_intra} demonstrate that extrapolating a link for a link loss beyond the training data leads to less accurate predictions. In contrast, Figure~\ref{fig:interpolation42} shows that when the test links' link losses are within the spectrum of the training links, the model provides better predictions. This underlines the importance of training data that fully represents the range of operational conditions.
This behavior shows that the ML model is better at interpolating than extrapolating.

%\newpage
\subsubsection*{Influence of Link Loss}

This section shows the influence of the link loss on the SKR predictions.
The prediction for link 5 was made with the before-mentioned interpolation model, and the prediction for link 4 was made with the extrapolation model.  
%The links that are show in Figure~\ref{fig:testpredlink45Linkloss} to demonstrate the effect of link loss on the SKR are...

\begin{figure}%[h]%[h!]
  \begin{center}
    \includegraphics[width=0.49\textwidth]{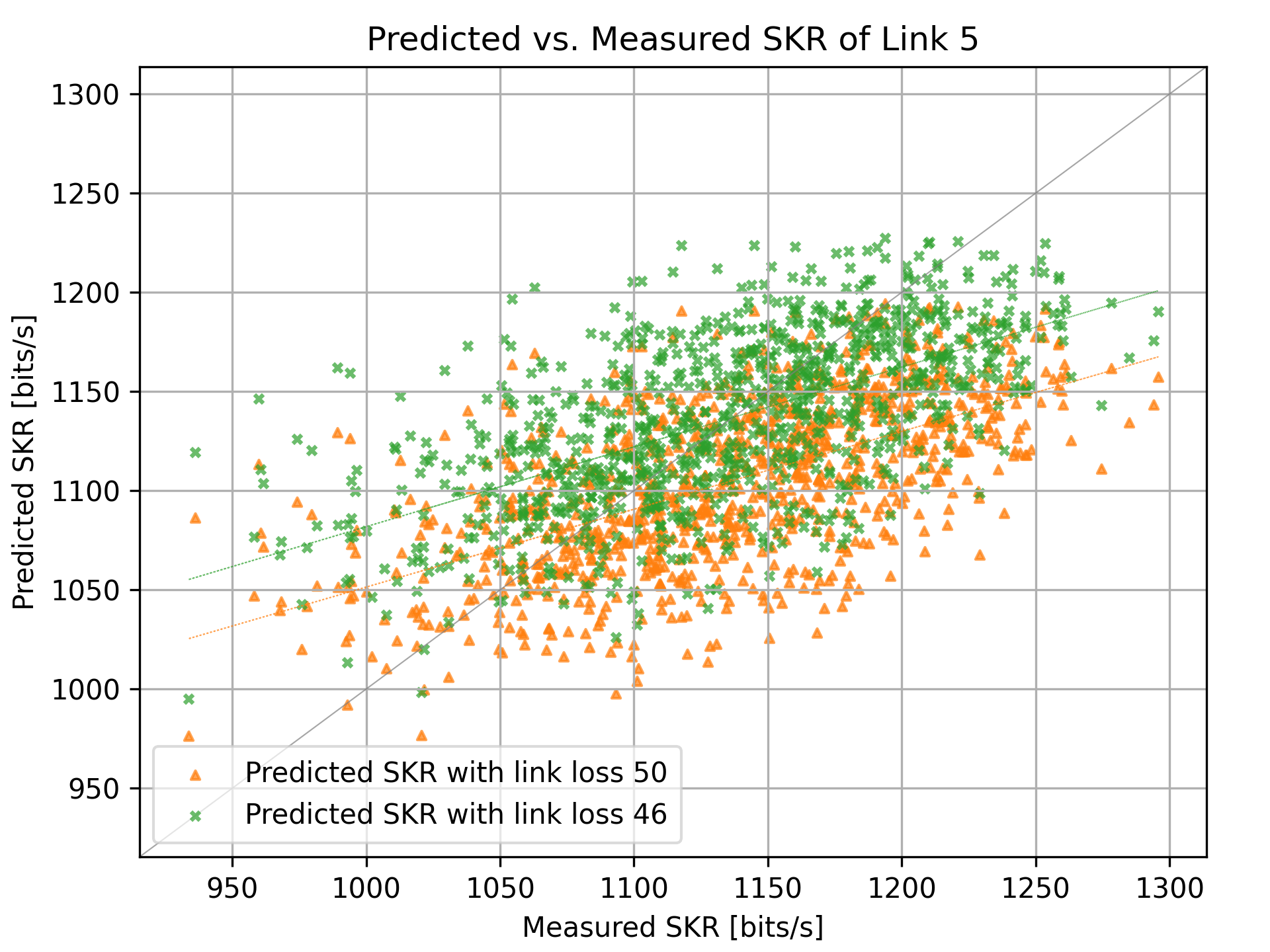}
    \includegraphics[width=0.49\textwidth]{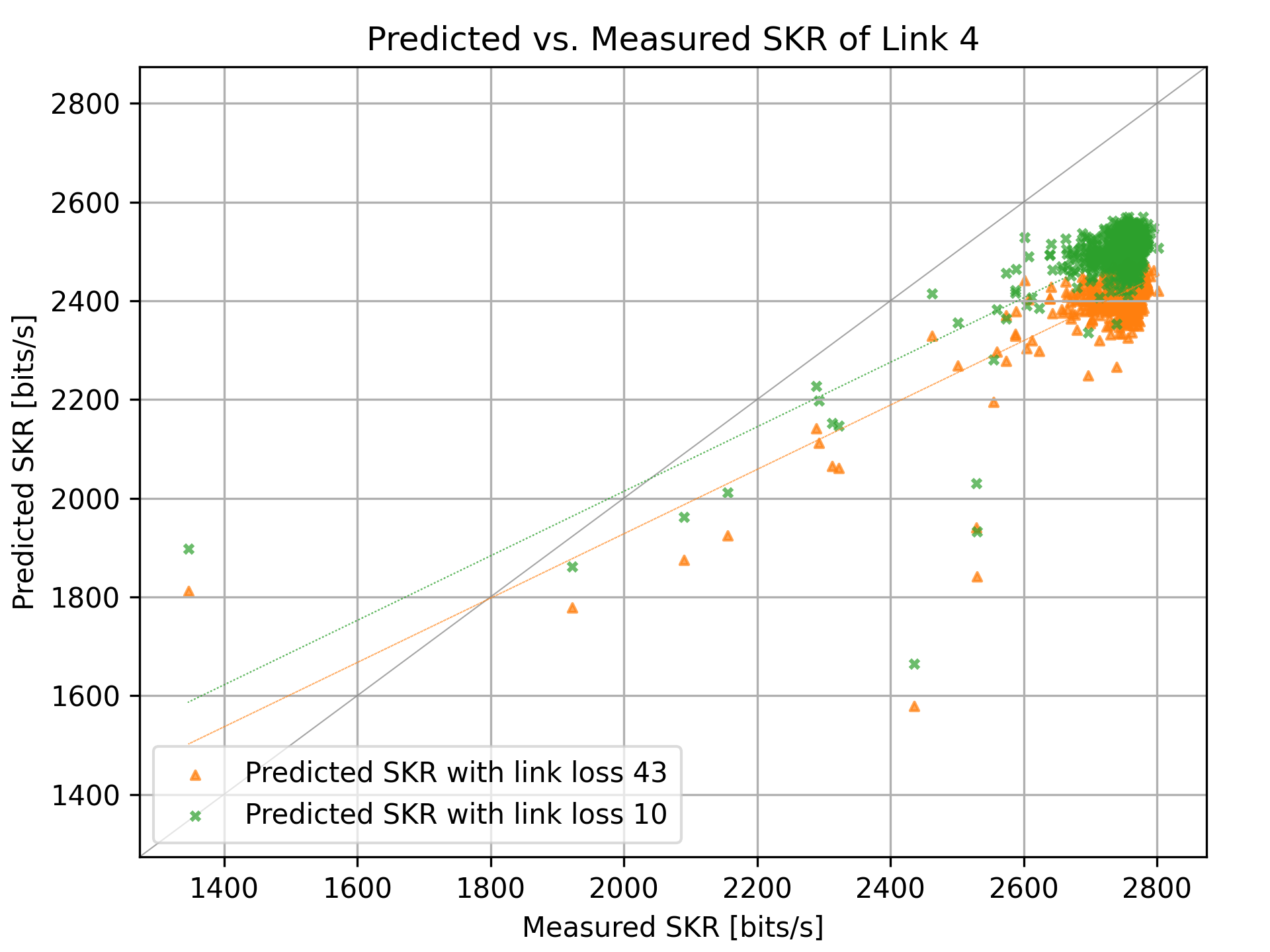}
    \caption{Comparing measured and predicted SKR for Link 4 and 5 with varying link loss parameters concerning the measured values}
    \label{fig:testpredlink45Linkloss}
  \end{center}
\end{figure}

\begin{table}%[h]
	\centering
	\begin{tabular}{|l|r|r|r|r|}
		\hline
		&
        \multicolumn{1}{c|}{ME} &
        \multicolumn{1}{c|}{MAE} &
        \multicolumn{1}{c|}{MRE} &
        \multicolumn{1}{c|}{MSE} \\
        \hline
		Link 5 with link loss 50	& -30.908906 & 48.440184 & -0.025329 & 0.000511 \\
 		\hline
		Link 5 with link loss 46 	& 0.884127 & 40.411869 & 0.002729 & 0.000376 \\
		\hline
		Link 4 with link loss 43	& -330.587135 & 331.537441 & -0.120235 & 0.016024 \\
		\hline
		Link 4 with link loss 20 	& -243.361182 & 244.48759 & -0.0884 & 0.008834 \\
		\hline
	\end{tabular}
	\caption{Comparing test errors of the interpolated and the extrapolated ML model}
	\label{tab:errors_linkloss}
\end{table}

Figure~\ref{fig:testpredlink45Linkloss} shows the predicted SKRs compared to the measured SKRs. The model used for prediction differs for both predicted curves where link 5 is interpolated  and link 4 is extrapolated. The legends indicate different link losses used for the model. Figure \ref{fig:testpredlink45Linkloss} suggests the influence of link loss on the SKR predicitons. 
Table~\ref{tab:errors_linkloss} shows the errors in the horizontal dimension and the links and models in the vertical dimension, demonstraiting the gained improvements in accuracy with varying link loss. 
Lowering link loss for both links leads to improved errors, as the predictor underestimated both links on the first run. However, link 4, which was extrapolated, does not show significant improvement for changing link losses. This my be due to extrapolation being challenging for the model, and the model struggling with link loss being outsite of the training range.

\newpage
\subsubsection*{Sophisticated ML Model Approach: Interpolation with 3 Training Links}

This interpolation approach aims to increase accuracy over the whole link loss spectrum covered by our training links. Three links are used as training data to reach this goal. This model uses links 1, 2 and 4 as training links with link losses of 46, 57 and 20 respectively.

\begin{figure}[h]%[h!]
  \begin{center}
    \includegraphics[width=0.90\textwidth]{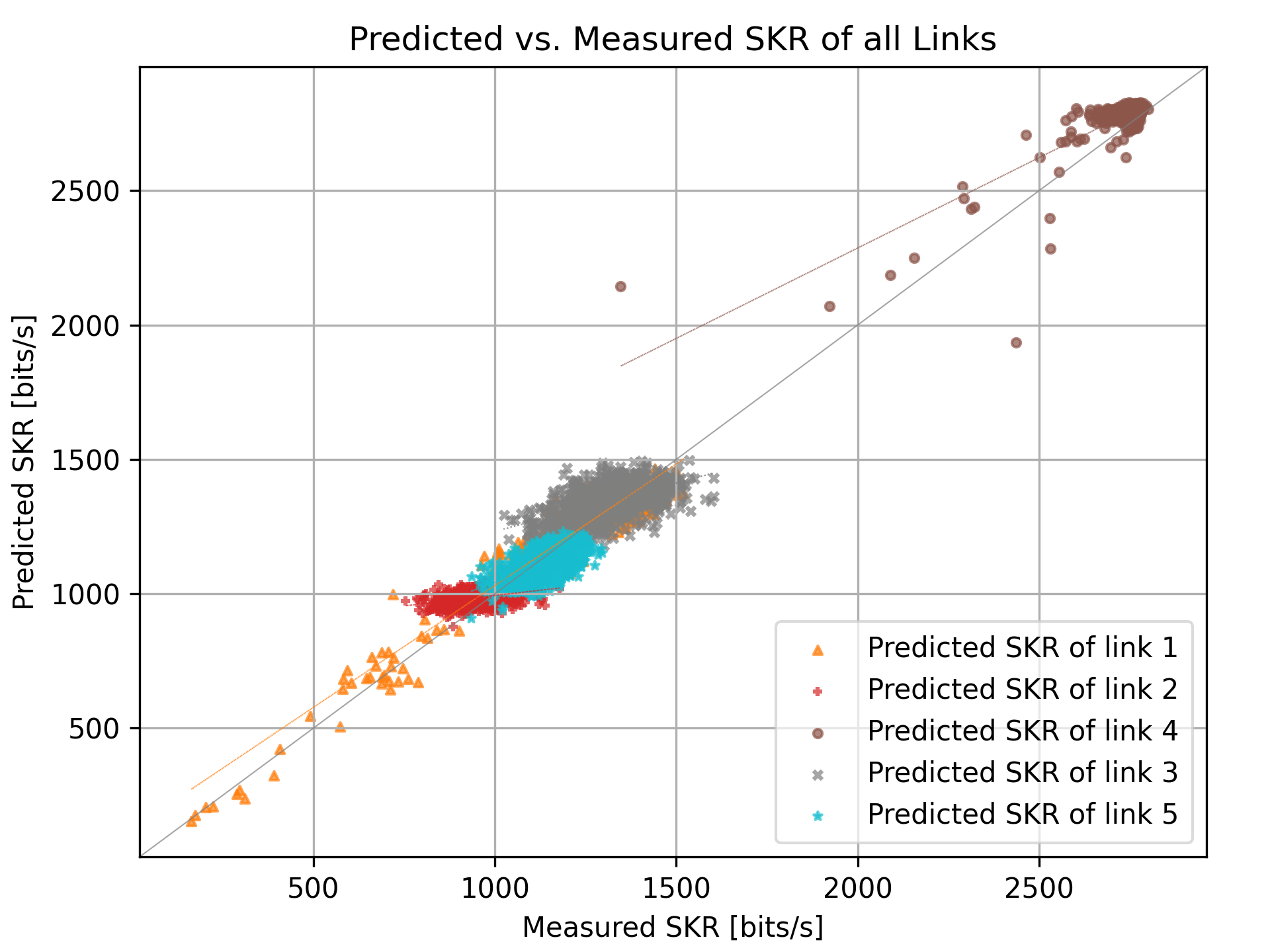}
    \caption{Comparing measured and predicted SKR of all train (1,2,4) and test (3,5) links of the final ML model}
    \label{fig:fullinterpolation}
  \end{center}
\end{figure}

\begin{table}%[h]
	\centering
	\begin{tabular}{|l|r|r|r|r|}
		\hline
		&
        \multicolumn{1}{c|}{ME} &
        \multicolumn{1}{c|}{MAE} &
        \multicolumn{1}{c|}{MRE} &
        \multicolumn{1}{c|}{MSE} \\
        \hline
		Link 1 (train) & 7.25297 & 43.68999 & 0.007575 & 0.000442 \\
		\hline
		Link 2 (train) & 22.2962 & 49.789192 & 0.027011 & 0.000573 \\
		\hline
		Link 3 (test) & 26.135826 & 68.076316 & 0.02335 & 0.001037 \\
		\hline
		Link 4 (train) & 43.163981 & 46.294002 & 0.01615 & 0.000532 \\
		\hline
		Link 5 (test) & -27.927293 & 46.865394 & -0.023011 & 0.000482 \\
		\hline
	\end{tabular}
	\caption{Comparing errors of all train (1,2,4) and test (3,5) links of the final ML model}
	\label{tab:errors_3_way_interpolation}
\end{table}

Figure~\ref{fig:fullinterpolation} shows the predicted SKRs compared to the measured SKRs. The same model is used for all scatter plots, but the legends indicate the different links that are being evaluated. Figure \ref{fig:fullinterpolation} gives an overview of all link predictions in one graph.
Table~\ref{tab:errors_3_way_interpolation} shows the errors in the horizontal dimension and the links in the vertical dimension. All links are marked if they are part of the train or test set. This table details the improvements in accuracy with one additional link in the training set compared to the former models. 
Figure~\ref{fig:fullinterpolation} shows how accurately all links are predicted. Table~\ref{tab:errors_3_way_interpolation} supports this observation showing lower test errors compared to Table~\ref{tab:errors_linkloss} where the error of link 5 is already improved before even varying the link loss. 

Therefore, we observe that this model generalizes better for various links with various link losses, fulfilling the aim of this more complex model.

%\subsection{Statistical Validation of Results}

%\section{Comparison and Validation}
%Comparison with Existing Models
%Validation Against Empirical Data

%  Conclusions (Zusammenfassung):
\chapter{Conclusions and Outlook}

%The thesis is concluded here. The considered problem is repeated. The contribution of this work is highlighted and the results are recapitulated. Remaining questions are stated and ideas for future work are expressed. 

%Answere to the questions from the Introduction

\section{Discussion of the Model's Accuracy and Practical Relevance}	%either this or a separate discussion chapter:

%\chapter{Discussion}

%Interpretation of Results
%Comparison with Existing Literature
%Theoretical and Practical Implications
%Limitations and Challenges

%The results have demonstrated that a theoretical COW model can translate theory into practice for forecasting the behavior of QKD systems, albeit with significant limitations. These limitations prompted us to adopt a machine learning (ML) approach. The ML model proved effective in forecasting QKD system behavior and successfully presented the corresponding data.

%In conclusion, our approach with the ML model confirms the potential for predicting the behavior of QKD systems in the future. This promising direction sets the stage for further research beyond the scope of this thesis, aiming to fully develop and refine these predictive capabilities.

This thesis aimed to explore how various environmental and operational parameters impact the performance  of QKD systems in real-world applications. Our research focused on how monitoring data can be effectively utilized to develop a physical layer model for QKD that accurately reflects these imperfections and predicts SKRs for various links. Therefore, we used input parameters such as QBER, visibility, and laser power to fit and train our models. We investigated key practical imperfections in QKD systems and how they deviate from idealized protocol conditions, leading to SKR fluctuation. Initially, we discussed various protocols, ultimately selecting and implementing a COW-based protocol for our theoretical calculations.

In the results we analyzed the effectiveness of a theoretical COW model and applied curve fitting to optimize its parameters. The results have shown that a theoretical COW model can translate theory into practice for forecasting the behavior of QKD systems, albeit with significant limitations. These limitations, such as the inability to detect abnormal system states and also the insufficient accuracy in the SKR estimations, prompted us to develop an ML approach. 

For the ML model approach, we discussed the minimization of our model, exploring how much we could reduce its size before losing accuracy. Conversely, we also optimized the model to achieve the most accurate SKR predictions.
Next, we incrementally fed the model with more input data and evaluated whether the error scaled with increasing inputs. Subsequently, we expanded the model to support training with two input links by introducing link loss (corresponding to link lengths) to differentiate between various links and predict new, unseen links. In this context, we discussed different prediction approaches: extrapolation versus interpolation. Finally, we developed a sophisticated model trained with three links, capable of minimizing test errors across all its test links. The ML model has proven effective in forecasting QKD system behavior more accurately than the COW model, demonstrating its potential by successfully estimating the SKR of unknown QKD links given only a few input parameters.

Our findings indicate that while the theoretical COW model provides a foundational understanding of QKD systems, the ML model is a more practical and accurate tool for predicting QKD system performance. By closely monitoring and analyzing environmental and operational parameters, the ML model can adapt to the complexities and imperfections present in real-world applications, thereby improving predictive accuracy. This confirms the potential for using ML models to forecast the behavior of QKD systems under various conditions, which is a crucial step towards enhancing the practical applicability and robustness of QKD systems.

%\subsection{Theoretical COW Model}

%This theoretical COW model proves that real-world systems can be represented theoretically.

\section{Outlook}

While our research demonstrates the ability to predict QKD system performance based on visibility and QBER, the challenge of forecasting future SKRs remains. Addressing the ability of predicting future SKRs with current data was beyond the scope of this thesis. Future research should focus on this area to further enhance the practical applicability and reliability of forecasting QKD systems behaviors in diverse scenarios.

In detail, developing a comprehensive physical layer model capable of long-term forecasting the SKR should be a priority for future research. Such a model could be instrumental in preventing encryption systems from running out of keys if the SKR drops significantly. Integrating these models into network management systems would allow for proactive adjustments, ensuring the continuous and secure operation of QKD networks. Additionally, understanding and predicting network-level behaviors, such as link stability and the impact of varying environmental conditions, could lead to the development of adaptive QKD systems that dynamically respond to network conditions. This would optimize not only prediction (capability) but also real-time performance, thereby enhancing reliability of future QKD networks.

In summary, this thesis establishes a foundational approach to using ML models for predicting and enhancing QKD system performance. Although significant progress has been made, further efforts are needed to fully realize the potential of ML models to predict QKD systems behaviors accurately.

%Improving a QKD system with a physical layer model that can adapt system parameters during runtime to optimize the transmission channel.

%-establisch a physical layer model that can not only predict data but also intervent into the QKD system and adapt parameters to optimize the transmission channel or even find a better channel for the secret key transmission.
%Looking forward, a key objective is to establish a comprehensive physical layer model capable of predicting data and actively intervening in QKD systems. This model would enable dynamic adaptation of system parameters to optimize the transmission channel, ensuring more reliable and efficient secret key generation. Additionally, such a model could facilitate the identification and selection of superior transmission channels, further enhancing the performance of QKD systems. While our current work does not implement this advanced physical layer model, we suggest that future research focuses on developing this capability to significantly improve the robustness and adaptability of QKD technology in practical applications.

%\include{Formatting}

% Appendix (Anhänge), could have multiple chaper-files:
\appendix
%\addcontentsline{toc}{chapter}{Appendix} % Manually add "Appendix" to the table of contents
\chapter{Technical Documentation}
%The appendix may contain some listings of source code that has been used for simulations, extensive proofs or any other things that are strongly related to the thesis but not of immediate interest to the reader.
This appendix provides comprehensive technical documentation relevant to the QKD system under study. It includes detailed specifications of the QKD system and examples of SKR data. The purpose of this appendix is to offer in-depth technical insights and empirical data that support the analysis and findings presented in the main body of this thesis. 

\newpage
\section{Cerberis\textsuperscript{3} QKD System}

\begin{table}[h!]
\centering
\begin{tabular}{|>{\raggedright\arraybackslash}p{0.40\textwidth}|>{\raggedright\arraybackslash}p{0.55\textwidth}|}
\hline
\textbf{Model} & \textbf{Cerberis\textsuperscript{3}} \\
\hline
\multicolumn{2}{|l|}{\textbf{KEY FEATURES}}\\
\hline
Key generation rate & 1.25GHz pulse repetition rate \\
\hline
High speed hardware-based key processing to distill the secret keys & yes \\
\hline
Key security parameter & $\epsilon_{QKD} = 4*10^{-9}$ \\
\hline
Dynamic range & 12 dB (up to 16/18 dB on request) \\
\hline
Maximum length of quantum channel (typ. @ 0.23dB/km) & 50 km (up to 70/80 km on request) \\
\hline
Secret key rate & 1.4kb/s (12dB) \\
\hline
\multicolumn{2}{|l|}{\textbf{PHYSICAL PARAMETERS}}\\
\hline
Dimensions & 19'' rackmount 6U ATCA chassis; 13'' depth \\
\hline
Switching Shelf Manager (SSM) & Full status LEDs, 10GbE SFP+, GbE ports, RS-232 de-bug ports, and Telco alarm \\
\hline
Power supply & Up to 4x swappable 1300W AC power supplies or 2x 90Amp DC Power Entry Modules. The input voltage is from 100 to 240 VAC or -36 to-72 VDC \\
\hline
Weight for one node & 30kg \\
\hline
\multicolumn{2}{|l|}{\textbf{Operating conditions}}\\
\hline
Temperature & 10 to 30°C \\
\hline
max. relative humidity (@30°C) & 80\% (non condensing) \\
\hline
\multicolumn{2}{|l|}{\textbf{Non-operating conditions}}\\
\hline
Temperature & -10 to +60°C \\
\hline
max. relative humidity (@40°C) & 90\% (non condensing) \\
\hline
\multicolumn{2}{|l|}{\textbf{MANAGEMENT AND MONITORING}} \\
\hline
Alerting functions & Temperature (high and low), fan failure, power supply failure, low key rate, high QBER (intruder alarm), service and key channel failure, key buffer \\
\hline
Continuous monitoring & Temperature, fan and power supply operations, system uptime, firmware version, CPU load, memory usage, actual quantum key rate, QBER, compression rate (due to key processing) \\
\hline
\end{tabular}
\caption{Specifications of the Cerberis\textsuperscript{3} QKD system from IDQ \cite{idq2019cerberis3}}
\label{tab:cerberis3}
\end{table}

\newpage
\section{SKR Data Example}

\subsection{SKR Raw Data}

\begin{table}[h]
\centering
\begin{tabular}{|>{\raggedright\arraybackslash}p{0.5\textwidth}|>{\raggedright\arraybackslash}p{0.4\textwidth}|}
\hline
\textbf{Timestamp} & \textbf{SKR} \\
\hline
2023-11-29 18:55:49.323195+00:00 & 1326 \\
\hline
2023-11-29 18:55:56.370954+00:00 & 1326 \\
\hline
2023-11-29 18:56:04.629459+00:00 & 1326 \\
\hline
2023-11-29 18:56:11.694552+00:00 & 1197 \\
\hline
2023-11-29 18:56:19.014720+00:00 & 1197 \\
\hline
2023-11-29 18:56:26.085442+00:00 & 1197 \\
\hline
2023-11-29 18:56:33.855276+00:00 & 1197 \\
\hline
2023-11-29 18:56:40.690756+00:00 & 1197 \\
\hline
2023-11-29 18:56:48.233272+00:00 & 1197 \\
\hline
2023-11-29 18:56:55.299210+00:00 & 1197 \\
\hline
2023-11-29 18:57:02.113707+00:00 & 1197 \\
\hline
2023-11-29 18:57:09.433454+00:00 & 1197 \\
\hline
2023-11-29 18:57:16.920912+00:00 & 1197 \\
\hline
2023-11-29 18:57:25.408257+00:00 & 1197 \\
\hline
2023-11-29 18:57:33.186220+00:00 & 1197 \\
\hline
2023-11-29 18:57:39.760193+00:00 & 1197 \\
\hline
2023-11-29 18:57:46.633200+00:00 & 1197 \\
\hline
2023-11-29 18:57:54.853191+00:00 & 1234 \\
\hline
2023-11-29 18:58:03.074196+00:00 & 1234 \\
\hline
\end{tabular}
\caption{SKR data from 2023-11-29}
\label{table:skrrawdata}
\end{table}

\subsection{SKR Filtered Data}

\begin{table}[h!]
\centering
\begin{tabular}{|>{\raggedright\arraybackslash}p{0.5\textwidth}|>{\raggedright\arraybackslash}p{0.4\textwidth}|}
\hline
\textbf{Timestamp} & \textbf{SKR} \\
\hline
2023-11-29 18:50:00+00:00 & 1222.53 \\
\hline
2023-11-29 19:00:00+00:00 & 1210.65 \\
\hline
2023-11-29 19:10:00+00:00 & 1284.35 \\
\hline
2023-11-29 19:20:00+00:00 & 1204.45 \\
\hline
2023-11-29 19:30:00+00:00 & 1155.12 \\
\hline
2023-11-29 19:40:00+00:00 & 1340.26 \\
\hline
2023-11-29 19:50:00+00:00 & 1286.22 \\
\hline
2023-11-29 20:00:00+00:00 & 1435.6 \\
\hline
\end{tabular}
\caption{SKR data from 2023-11-29}
\label{tab:skrfiltereddata}
\end{table}

% Abbreviations (Abkürzungsverzeichnis):
%\chapter{Notation und Abkürzungen}
\chapter{Abbreviations and Terminology}
This appendix contains tables of all abbreviations and other notations like mathematical
placeholders that were used in the thesis.
\begin{table}[h]
\begin{tabular}{ll}
AES		&	Advanced Encryption Standard\\
Alice 	& 	Represents the sender in cryptographic protocols\\
AU (a.u.)&	Arbitrary unit (relative unit of measurement)\\
Bob 	& 	Represents the receiver in cryptographic protocols\\
COW		&	Coherent One-Way\\
ES		&	Eigenstate\\
Eve 	& 	Represents an eavesdropper or adversary in cryptographic protocols\\
CSV		&	Comma Separated Values\\
FFNN	& 	Feedforward Neural Network\\
$I_{ab}$	&	Information shared betreen Alice and Bob\\
$I_{ae}$	&	Information shared betreen Alice and Eve\\
IDQ		&	ID Quantique\\
MAE		&	Mean Absolute Error\\
MB		&	Measurement Basis\\
ME		&	Mean Error\\
ML		& 	Machine Learning\\
MRE		&	Mean Relative Error\\
MSE		&	Mean Squared Error\\
ONB		&	Orthonormal Basis\\
OTP		&	One-Time-Pad\\
PNC		&	Photon-Number-Counting\\
PNS		&	Photon Number Splitting\\
QKD 	&	Quantum Key Distribution\\
ReLU	&	Rectified Linear Unit\\
RNN 	&	Recurrent Neural Network\\
RSA		&	Rivest–Shamir–Adleman\\
SKR 	&	Secret Key Rate\\
SSPD	&	superconducting single-photon detector\\
SNSPD	&	superconducting nanowire single-photon detector\\
\end{tabular}
\end{table}

% References (Literaturverzeichnis):
% a) Style (with numbers: use unsrt, for author+year: use alpha):
%\bibliographystyle{unsrt}
\bibliographystyle{IEEEtran}
% b) The File:
\bibliography{Bibliography}
%if bib is not shown, compile Document in the following order: 1. PDFLaTex 2. BibTeX 3. 2x PDFLaTex

%%%%%%%%%%%%%%%%%%%%%%%%%%%%%%%%%%%%%%%%%%%%%%%%%%%%%%%%%%%%

%%%%%%%%%%%%%%%%%%%%%%%%%%%%%%%%%%%%%%%%%%%%%%%%%%%%%%%%%%%%

%%%%%%%%%%%%%%%%%%%%%%%%%%%%%%%%%%%%%%%%%%%%%%%%%%%%%%%%%%%%
\end{document}